\documentclass[a4paper,12pt, epsfig]{article}
\def\letter{0}\def\pr{0}
\pdfoutput=1 
\usepackage{epsfig}
\usepackage{epstopdf}
\usepackage{graphicx}
\usepackage{ifthen}
\usepackage{ulem}

\usepackage{amsmath}
\usepackage[psamsfonts]{amssymb}
\usepackage{euscript}

\usepackage{latexsym}
\usepackage[arrow,matrix,curve]{xy}

\usepackage[hypertexnames=false]{hyperref}
\hypersetup{
    colorlinks=true,
    linkcolor=blue,
    filecolor=magenta,   
    citecolor=black,   
    urlcolor=cyan,
    }

\newskip\humongous \humongous=0pt plus 1000pt minus 1000pt

\newif\ifdtup

\def\,{\hspace{-.1cm}}
\def\hsp{,\hspace{.7cm}}

\def\fc#1#2 {\frac{n}{q}#1\frac{n}{q}#2}

\def\kt{\kappa}
\def\kt{\mathfrak{K}}
\def\ks{|\kt\rangle}

\def\rv{{\rm{vac}}}

\newcommand{\vac}{\ensuremath{|0\rangle}}

\renewcommand{\sin}{\textrm{sin}}

\renewcommand{\sinh}{\textrm{sinh}}

\renewcommand{\tanh}{\textrm{tanh}}
\newcommand{\sech}{\textrm{sech}}
\newcommand{\csch}{\textrm{csch}}

\def\sign#1{{\rm sign}\left(#1\right)}

\renewcommand{\theequation}{\arabic{section}.\arabic{equation}}
\renewcommand{\(}{\begin{equation}}
\renewcommand{\)}{end{equation} \vspace{-.05in}\linebreak}

\newcounter{saveeqn}
\newcounter{savealpheqn}

\newcommand{\alpheqn}{\setcounter{saveeqn}{\value{equation}}%
  \stepcounter{saveeqn}\setcounter{equation}{0}%
  \renewcommand{\theequation}{\mbox{\arabic{section}.\arabic{saveeqn}
\alph{equation}}}
  \renewcommand{\)}{\end{equation}}}
\def\part#1{\frac{\partial}{\partial{#1}}}%
\def\group#1{\refstepcounter{equation}\setcounter{saveeqn}
 {\value{equation}}%
  \label{#1}\setcounter{equation}{0}%
\renewcommand{\theequation}{\mbox{\arabic{section}.\arabic{saveeqn}
\alph{equation}}}
  \renewcommand{\)}{\end{equation}}}
\newcommand{\reseteqn}{\setcounter{equation}{\value{saveeqn}}%
  \renewcommand{\theequation}{\arabic{section}.\arabic{equation}}%
  \renewcommand{\)}{\end{equation}}}

\newcommand{\aalpheqn}{\setcounter{saveeqn}{\value{equation}}%
  \stepcounter{saveeqn}\setcounter{equation}{0}%
  \renewcommand{\theequation}{\mbox{
        \Alph{subsection}.\arabic{saveeqn}\alph{equation}}}
   \renewcommand{\)}{\end{equation}}}
\newcommand{\areseteqn}{\setcounter{equation}{\value{saveeqn}}%
  \renewcommand{\theequation}{\Alph{subsection}.\arabic{equation}}%
  \renewcommand{\)}{\end{equation}}}

\renewcommand{\thefootnote}{\alph{footnote}}
\renewcommand{\(}{\begin{equation}}
\renewcommand{\)}{\end{equation}}
\newcommand{\ba}{\begin{eqnarray}}
\newcommand{\ea}{\end{eqnarray}}

\renewcommand{\b}{\beta}

\renewcommand{\sl}{{g}}

\newcommand{\cbp}{\mathop{\vtop{\ialign{##\crcr
   $\hfil\displaystyle{}\hfil$\crcr\noalign{\kern-13pt\nointerlineskip}
   \BIG{)}\hskip0pt\crcr\noalign{\kern3pt}}}}}
\newcommand{\pa}{\mathop{\vtop{\ialign{##\crcr

$\hfil\displaystyle{\oplus}\hfil$\crcr\noalign{\kern+1pt\nointerlineskip
}
   \hspace{.08in}$^{\alpha=0}$\hskip6pt\crcr\noalign{\kern3pt}}}}}
\renewcommand{\hsp}{,\hspace{.3in}}
\newcommand{\p}{^\prime}
\newcommand{\pp}{^{\prime\prime}}

\catcode`\@=11
\def\vereq#1#2{\lower3pt\vbox{\baselineskip1.5pt \lineskip1.5pt
\ialign{$\m@th#1\hfill##\hfil$\crcr#2\crcr\sim\crcr}}}
\catcode`\@=12

\renewcommand{\(}{\begin{equation}}
\renewcommand{\)}{\end{equation}}

\def\pin#1{\int \frac{d#1}{2\pi}}
\def\ppin#1{\int\hspace{-17pt}\sum \frac{d#1}{2\pi}}
\def\ppink#1{\int\hspace{-17pt}\sum\frac{d^{#1}k}{(2\pi)^{#1}}}
\def\ppinkp#1{\int\hspace{-17pt}\sum\frac{d^{#1}k\p}{(2\pi)^{#1}}}
\def\dint{\int\hspace{-12pt}\sum }
\def\pink#1{\int \frac{d^{#1}k}{(2\pi)^{#1}}}

\def\sQ{{\sqrt{Q_0}}}

\def\Bd#1{B^\ddag_{k_{#1}}}

\def\cc{\mathcal{C}}
\def\df{\mathcal{D}_{f}}

\def\L{{\rm{(L)}}}
\def\I{\mathcal{I}}

\def\os{\omega_S}

\def\blu#1{\textcolor{blue}{Jarah: #1}}
\def\red#1{\textcolor{red}{Hui: #1}}

\newcommand{\beas}{\begin{eqnarray*}}
\newcommand{\eeas}{\end{eqnarray*}}

\newcommand{\bquo}{\begin{quote}}
\newcommand{\enqu}{\end{quote}}

\def\lim#1{\stackrel{\rm{lim}}{{}_{#1}}}

\newcommand{\R}{{\mathbb R}}

\newcommand{\g}{\mathfrak g}

\def\ch{{\mathcal{H}}}

\def\ok#1{\omega_{k_{#1}}}
\def\okp#1{\omega_{k\p_{#1}}}
\def\okt#1{\omega_{\kt_{#1}}}
\def\V#1{V^{(#1)}(gf(x))}
\def\v#1{V^{(#1)}[f(x),x]}
\def\ck{\csch\left(\frac{\pi k}{2\b}\right)}

\def\mb{\mathcal{B}}
\def\mc{\mathcal{C}}
\def\md{\mathcal{D}}
\def\me{\mathcal{E}}

\newcommand{\beq}{\begin{equation}}
\newcommand{\eeq}{\end{equation}}
\newcommand{\bea}{\begin{eqnarray}}
\newcommand{\eea}{\end{eqnarray}}

\newskip\humongous \humongous=0pt plus 1000pt minus 1000pt

\newif\ifdtup

\catcode`\@=11

\ifthenelse{\equal{\letter}{0}}{ 

\@addtoreset{equation}{section}
\def\theequation{\arabic{section}.\arabic{equation}}

\def\@normalsize{\@setsize\normalsize{15pt}\xiipt\@xiipt
\abovedisplayskip 14pt plus3pt minus3pt%
\belowdisplayskip \abovedisplayskip
\abovedisplayshortskip \z@ plus3pt%
\belowdisplayshortskip 7pt plus3.5pt minus0pt}

\def\small{\@setsize\small{13.6pt}\xipt\@xipt
\abovedisplayskip 13pt plus3pt minus3pt%
\belowdisplayskip \abovedisplayskip
\abovedisplayshortskip \z@ plus3pt%
\belowdisplayshortskip 7pt plus3.5pt minus0pt
\def\@listi{\parsep 4.5pt plus 2pt minus 1pt
      \itemsep \parsep
      \topsep 9pt plus 3pt minus 3pt}}

\relax

\def\section{\@startsection{section}{1}{\z@}{3.5ex plus 1ex minus  .2ex}{2.3ex plus .2ex}{\large\bf}}

\def\thesection{\arabic{section}}
\def\thesubsection{\arabic{section}.\arabic{subsection}}

\def\appendix{\setcounter{section}{0}
 \def\thesection{Appendix \Alph{section}}
 \def\thesubsection{\Alph{section}.\arabic{subsection}}
 \def\theequation{\Alph{section}.\arabic{equation}}}
\renewcommand{\theequation}{\arabic{section}.\arabic{equation}}

}{
\renewcommand{\theequation}{\arabic{equation}}

} 

\begin{document}
\def\thefootnote{\fnsymbol{footnote}}
\def\thetitle{Asymptotic States for Kink-Meson Scattering}
\def\auttwo{Hui Liu}
\def\autone{Jarah Evslin}
\def\affc{School of Fundamental Physics and Mathematical Sciences, Hangzhou Institute for Advanced Study,
University of Chinese Academy of Sciences, Hangzhou 310024, China}
\def\affb{University of the Chinese Academy of Sciences, YuQuanLu 19A, Beijing 100049, China}
\def\affd{Arnold Sommerfeld Center, Ludwig-Maximilians-Universität, Theresienstraße 37, 80333 München, Germany}
\def\affa{Institute of Modern Physics, NanChangLu 509, Lanzhou 730000, China}
\def\affe{Institute of Contemporary Mathematics, School of Mathematics and Statistics,
Henan University, Kaifeng, Henan 475004, P. R. China}


\ifthenelse{\equal{\pr}{1}}{
\title{\thetitle}
\author{\autone}
\author{\auttwo}
\author{\autthree}
\affiliation {\affa}
\affiliation {\affb}

}{

\begin{center}
{\large {\bf \thetitle}}

\bigskip

\bigskip


{\large \noindent  \autone{${}^{1,2}$} \footnote{jarah@impcas.ac.cn} 
}


\vskip.7cm

1) \affa\\
2) \affb\\

\end{center}

}

\begin{abstract}
\noindent
The definition of a quantum state corresponding to a wave packet far from a global soliton is considered.  We define an asymptotic quantum state corresponding to a localized wave packet of elementary quanta far from a kink.  We demand that the state satisfies two properties.  First, it must evolve in time via a rigid translation of the wave packet, up to the usual wave packet spreading and corrections which are exponentially suppressed in the distance to the kink.  Second, the state must be invariant under a simultaneous translation of the kink and the wave packet.  
We explicitly construct the leading quantum corrections to an asymptotic state consisting of a meson approaching a kink.  
We expect this construction to readily generalize to elementary quanta in the presence of any global soliton.

\end{abstract}

%
\setcounter{footnote}{0}
\renewcommand{\thefootnote}{\arabic{footnote}}

\ifthenelse{\equal{\pr}{1}}
{
\maketitle
}{}

\section{Introduction} \label{insez}

\subsection{Nonrelativistic Quantum Mechanics}

Consider the scattering off of a localized potential in nonrelativistic quantum mechanics.  There are two ways to compute the reflection coefficient.  First, one may begin with a localized wave packet
\beq
\psi(t=0,x)=\psi_{x_0}(x)
\eeq
arriving from the far left $x_0\ll 0$.  One evolves the system using the time-dependent Schrodinger equation
\beq
\psi(t,x)=e^{-iHt}\psi(t=0,x).
\eeq
Then, at some sufficiently large time, one takes the inner product of the wave packet with a basis of outgoing wave packets on the far left.  A linear combination of these inner products is the reflection coefficient.  The  second approach is to begin with a non-localized Hamiltonian eigenstate which has no incoming part on the far right.  One then directly computes the inner product of this eigenstate with the far left, outgoing wave packets to obtain the reflection coefficient.

These two approaches both require one to define asymptotic, localized wave packets.  In the first case, one is needed for the initial condition and one for the basis of final states.  In the second case, only one is needed, for the final states.  These asymptotic states are necessary to calculate the reflection coefficient.  They are not determined by the Hamiltonian eigenstates alone.  The asymptotic state is not a Hamiltonian eigenstate, which would not evolve.  However, up to the usual wave packet spreading effects, which vanish in the monochromatic limit, for some shape $\psi(x)$ and velocity $v$ they satisfy
\beq
\psi(t,x)=\psi_{x_t}(x)\hsp x_t=x_0+vt\hsp \psi_{y}(x)=\psi(x-y). \label{rigid}
\eeq

\subsection{Kink-Meson Scattering}

Kink-kink and kink-antikink scattering are known to be rich subjects classically \cite{csw,doreyf6,sfal21,multex22a,multex22b,alonso22,col22} and quantum mechanically \cite{tanmay21,takyi22,tanmay23}.  However the scattering of kinks with elementary mesons has received relatively little attention even classically \cite{tomrad1,tomrad2,tomrad3} let alone in quantum field theory \cite{faddeev77,lowe79,parm87,swanson88,uehara91,hayashi1,hayashi2,abdel11}.  Recently, we have begun a systematic treatment of such processes, as an expansion in the coupling $g$.  The order $O(g^0)$ scattering, corresponding to classical wave mechanics, was treated in Ref.~\cite{merefl}.  

Next an exhaustive treatment of scattering processes with probabilities of order $O(g^2)$ was performed.  There are only three processes, meson multiplication \cite{memult}, in which the final state consists of two mesons and a kink, and (anti)-Stokes scattering \cite{mestokes}, in which the final state consists of a single meson and a kink but an internal kink excitation is toggled.  Meson multiplication was treated by evolving an incoming wave packet in Ref.~\cite{memult} and by beginning with a full Hamiltonian eigenstate in Ref.~\cite{menorm}, reflecting the two strategies in nonrelativistic quantum mechanics described above.

In all of these studies, asymptotic wave packets were required.  However, for these leading order processes, the quantum corrections to the asymptotic states were not important.  Nonetheless, to push our study to $O(g^4)$, where elastic meson-kink scattering appears for classically reflectionless kinks, quantum corrections to the initial and final states are important.  We expect such elastic scattering to be phenomenologically rich, as, for example, one may expect to discover an infinite tower of narrow resonances in the reflection probability corresponding to multiply excited shape modes.

\subsection{Defining Asymptotic States}

This is the motivation for the current work.  We seek to define the quantum corrections to the asymptotic states needed to consider kink-meson scattering.  Of course, the choice of asymptotic state is a choice, dictated by the experimental conditions.   Intuitively, we want the incoming asymptotic meson wave packet to not know about the kink.  Concretely, we will define our choice to be that which satisfies two criteria.  First,  similarly to Eq.~(\ref{rigid}), the wave packet should be characterized by a position and, when the wave packet is far from the kink, time evolution should only change this position, up to the usual spreading effects which disappear in the monochromatic limit.  This is rather nontrivial in quantum field theory, as the state corresponding to a single particle will contain, for example, a superposition of pairs of off-shell particles that naively travel at different speeds.  This condition will therefore strongly constrain the quantum corrections.  

The second condition that we will impose is that the state is annihilated by the momentum operator.  This operator simultaneously translates the mesons and the kink.  So this leads to a superposition of states where the kink position is different, indeed anywhere, but the kink-meson distance is fixed.  This condition is necessary if we are to use the reduced inner product in Ref.~\cite{menorm} to evaluate matrix elements. 

\subsection{Constructing Asymptotic States}

Of course, the definition of asymptotic states in the scattering of ordinary quanta is quite standard.  One simply calculates the Hamiltonian eigenstate corresponding to a single particle, which is an infinite sum over various $n$-particle Fock states, and folds it into a wave packet.  Then two such wave packets, which are well-separated, are tensored together.  This is a reasonable procedure, as the two particles do not interact.

In the case of a kink and a meson, or more generally any particle and a domain wall or even an arbitrary global soliton such as a Skyrmion \cite{skyrme,smorg}, the above, standard procedure does not apply.  In a sense the kink and the meson do not interact at a distance, as they do not exert a force on one another.  Nonetheless, the presence of the kink profoundly influences the elementary meson, determining, for example, the expectation value of the scalar field in the neighborhood of the meson.  Thus one cannot simply consider the kink and meson in isolation and tensor them, as a meson in isolation makes no sense without specifying the local vacuum, which depends on its relative position with respect to the kink.

Our approach will therefore be quite different.  We will use the linearized perturbation theory of Refs.~\cite{mekink,me2loop}.  Here one writes all states and operators in the kink frame, which is related to the usual frame by a unitary transformation that shifts the field by the kink solution.  The Hamiltonian in this frame is called the kink Hamiltonian.  States are decomposed into graded components depending on the number of zero modes, which translate the kink.  The gradings are nonnegative integers.  We refer to the zero-grading part of a state as the primary part and the rest as the descendants.

The construction will be as follows.  First, all descendants will be fixed by demanding exact translation-invariance.  This is sufficient for translation invariance to be exact, satisfying our second criterion.  Then the primary part will be fixed by a kind of Hamiltonian eigenvalue equation.  However, the state will be separated into a dressed meson operator acting on a dressed kink state, where each dressing corresponds to a cloud of virtual mesons.  We keep track of which meson is in which cloud.  Then we act the kink Hamiltonian on the dressed kink, but we instead introduce a vacuum Hamiltonian which is applied to the dressed meson.  The vacuum Hamiltonian is constructed from the kink Hamiltonian by taking the limit that the distance to the kink is infinite, before any other limit is taken. 

\subsection{Summary}

After a long introduction in Sec.~\ref{insez}, we review linearized soliton perturbation theory in Sec.~\ref{revsez}.  Our general construction of asymptotic states will be presented in Sec.~\ref{princsez}.  Next we find the leading corrections to a state with a single meson and a single kink in Sec.~\ref{gsez}.  This result is Eq.~(\ref{init}).  It justifies some steps that were simply asserted in Refs.~\cite{memult,menorm}.  In Sec.~\ref{movesez} we show that indeed the meson wave packet moves rigidly, despite the fact that the vacuum Hamiltonian was used in its construction, which does not commute with the translation operator.  The terms resulting from this failed commutation vanish when folded into the wave packet.  The next order correction, which is in principle necessary for $O(g^4)$ kink-meson scattering, is computed in Sec.~\ref{ggsez}.  This allows us to fix the parameter in our definition of the asymptotic state which corresponds to the eigenvalue in the usual eigenvalue problem.  It turns out to be the sum of the energy of the ground state kink with the energies of the mesons in the vacuum in which their wave packet is localized, and apparently agrees with the total energy.  

\section{Review} \label{revsez}

The definition of an asymptotic state for an incoming wave packet is nontrivial in the presence of any global soliton, be it a Skyrmion, an extended domain wall or simply a kink, because the soliton affects the choice of vacuum around the wave packet.  We believe that the procedure described in this note for defining such an asymptotic state can be straightforwardly applied in any of these contexts.   

However, we will specialize in this paper to the case of a scalar field theory in 1+1 dimensions.  This will make the discussion more concrete and simplify matters, as all ultraviolet divergences may be removed by the usual normal ordering $::_a$ and also linearized kink perturbation theory is available, which greatly simplifies computations in the one-kink sector.  In this section we will review linearized perturbation theory, as developed in Refs.~\cite{mekink,me2loop}.

\subsection{The Classical Field Theory}

For simplicity we consider a single scalar field $\phi(x)$ and its conjugate momentum $\pi(x)$, described by the Hamiltonian
\beq
H=\int dx :\ch(x):_a\hsp
\ch(x)=\frac{\pi^2(x)}{2}+\frac{\left(\partial_x \phi(x)\right)^2}{2}+\frac{V(g \phi(x))}{g^2}. \label{ham}
\eeq
We will consider a perturbative expansion in $g$, or equivalently a semiclassical expansion in the dimensionless quantity $g\sqrt{\hbar}$.  We are interested in potentials $V$ with two degenerate minima, so that there are classical kink solutions $\phi(x,t)=f(x)$ which interpolate between the minima.  We choose a single such solution $f(x)$, breaking the manifest translation-invariance.  Below we will describe how translation-invariant states may nonetheless be constructed.

We further demand that the second derivatives of the potential $V$ agree at the two minima
\beq
m^2=V^{(2)}_{\pm}\hsp
V^{(n)}_{\pm}=V^{(n)}(\sl f(\pm\infty))
\eeq
where $V^{(n)}$ is the $n$th derivative of $V(g\phi(x))$ with respect to its argument $g\phi(x)$.  If they do not agree, then the one-loop corrections to the vacuum energies on the two sides of the kink will not agree, and the kink will be a false vacuum bubble wall which will accelerate \cite{wstabile}.  We are interested in kink states which are eigenstates of the Hamiltonian, and so will not be interested in such accelerating solutions. 

Our perturbative expansion will correspond to small perturbations about the kink solution.  Such perturbations may be decomposed into normal modes $\g(x)$ of frequency $\omega$ which satisfy the Sturm-Liouville equation
\beq
\V{2}{\g}(x)=\omega^2{\g}(x)+{\g}^{\prime\prime}(x)\hsp \phi(x,t)=f(x)+e^{-i\omega t}\g(x). \label{sl}
\eeq
There is always a single zero mode $\g_B(x)$ with $\omega_B=0$.  There may be discrete shape modes $\g_S(x)$ with $0<\os<m$, which we take to be real.  Finally, there are continuum modes $\g_k(x)$ for all real $k$ with
\beq
\ok{}=\sqrt{m^2+k^2}.
\eeq
Although the generalization is straightforward \cite{merefl}, in this paper we consider classically reflectionless kinks, such as those in the Sine-Gordon and $\phi^4$ double-well models, for concreteness.  For these, we fix the sign of $k$ such that $\g^*_k(x)=\g_{-k}(x)$ and, for $|x|\gg 0$, $\g_k(x)$ is proportional to $e^{-ikx}$.  We fix the sign of $\g_B$ via
\beq
\g_B(x)=-\frac{f\p(x)}{\sqrt{Q_0}} \label{gb}
\eeq
where $Q_0$ is the energy of the classical kink.  When we pass to the quantum theory, we will define $Q_i$ to be the order $O(g^{2i-2})$ correction to the energy of the kink ground state.

As Eq.~(\ref{sl}) is of Sturm-Liouville type, the normal modes are a basis of bounded functions.  As a result, the normalization conditions
\beq
\int dx |{\g}_{B}(x)|^2=1,\
\int dx {\g}_{k_1} (x) {\g}^*_{k_2}(x)=2\pi \delta(k_1-k_2),\ 
\int dx {\g}_{S_1}(x){\g}^*_{S_2}(x)=\delta_{S_1S_2} \label{comp}
\eeq
lead to the completeness relation
\beq
\g_B(x)\g_B(y)+\ppin{k} \g_k(x)\g_{-k}(y)=\delta(x-y)\hsp
\ppin{k}=\pin{k}+\sum_S.
\eeq

\subsection{Quantization}

We will quantize this field theory in the Schrodinger picture by imposing the canonical commutation relations
\beq
[\phi(x),\pi(y)]=i\delta(x-y). \label{ccr}
\eeq
This paper will be entirely in the Schrodinger picture.  As the normal modes are a basis of functions of $x$, and Schrodinger picture operators are independent of $t$, we may decompose them in terms of normal modes \cite{cahill76}
\bea
\phi(x) &=&\phi_0 \mathfrak{g}_B(x)+\ppin{k} \left(B_k^{\ddag}+\frac{B_{-k}}{2 \omega_k}\right) \mathfrak{g}_k(x)\hsp
B^\ddag_k=\frac{B^\dag_k}{2\ok{}}\hsp
B^\ddag_S=\frac{B^\dag_S}{2\omega_S} \label{dec}\\
\pi(x) &=&\pi_0 \mathfrak{g}_B(x)+i \ppin{k}\left(\omega_k B_k^{\ddag}-\frac{B_{-k}}{2}\right) \mathfrak{g}_k(x)\hsp
B_S=B_{-S}.  \nonumber
\eea
This decomposition is invertible, and so one can use Eq.~(\ref{ccr}) to show that the basis of operators $\phi_0,\ \pi_0,\ B^\ddag_S,\ B_S,\ \Bd{}$\ and $B_{k}$ satisfies the algebra 
\beq
\left[\phi_0, \pi_0\right]=i, \quad\left[B_{S_1}, B_{S_2}^{\ddagger}\right]=\delta_{S_1 S_2}, \quad\left[B_{k_1}, B_{k_2}^{\ddagger}\right]=2 \pi \delta\left(k_1-k_2\right). \label{cr}
\eeq
We will see in Eq.~(\ref{pos}) that the zero-mode operator $\phi_0$ is proportional to the kink position, and $\pi_0$ to its momentum.  Note that $\phi_0$ only agrees with the collective coordinate of Refs.~\cite{cl75,gjscc} at linear order, as $\g_B(x)$ in Eq.~(\ref{gb}) implies that $f(x)+\phi_0\g_B(x)$ is a linear truncation of the shifted kink solution $f(x-\phi_0/\sQ)$.  $B^\ddag_S$ and $\Bd{}$ create perturbative shape modes and continuum modes respectively, while $B_S$ and $B_{k}$ destroy them. 

We have defined the Hamiltonian in terms of the usual normal ordering $::_a$ in which operators are expanded in plane waves with coefficients $a^\dag$ and $a$, and all $a^\dag$ are placed on the left.  We will call this plane-wave normal ordering.  However, all operators can also be written in the basis $\phi_0,\ \pi_0,\ B^\ddag_S,\ B_S,\ \Bd{}$\ and $B_{k}$.  In this basis, another normal ordering prescription, called normal-mode normal ordering, will be more convenient.  This is defined by placing all $B$ and $\pi_0$ on the right.  We will denote this prescription by $::_b$.  The two normal ordering prescriptions are related by a Wick's theorem \cite{mewick}
\bea
:\phi^j(x):_a&=&\sum_{m=0}^{\lfloor{\frac{j}{2}}\rfloor}\frac{j!}{2^m m!(j-2m)!}\I^m(x):\phi^{j-2m}(x):_b\label{wick}\\
\I(x)&=&\pin{k}\frac{\left|\g_{k}(x)\right|^2-1}{2\omega_k}+\sum_S\frac{\left|{\g}_{S}(x)\right|^2}{2\omega_S}.\nonumber
\eea

We will always consider asymptotic expansions about zero coupling.  Here the Hilbert space of states can be decomposed into sectors corresponding to different numbers of kinks.  The vacuum lies in the zero-kink sector.  More precisely, there is a vacuum sector for each minimum of $V$.  In this paper we will be interested in the one-kink sector.  States in this sector consist of a single kink plus the Fock space of a finite number of perturbative excitations, which we call mesons.  Sometimes we will distinguish excitations which are bound to the kink from continuum excitations, referring to the bound excitations as shape modes.

\subsection{The Kink Frame}

The standard perturbative expansion treats the higher powers of $\phi$ in the Hamiltonian as small perturbations.  This is reasonable if $\phi$ is in some sense small.  In classical field theory, it is an expansion about $\phi=0$.  In quantum field theory, correspondingly, it can only provide a reasonable approximation if the expectation values of powers of $\phi$ are close to zero.  

On the other hand, in the one-kink sector, the expectation value of $\phi$ is equal to $f(x)$ plus corrections suppressed by a power of $g$.  We are therefore interested in an expansion about $\phi(x)=f(x)$.  In classical field theory, this would be easy to arrange.  One would simply define $\eta(x,t)=\phi(x,t)-f(x)$ and perturbatively treat $\eta(x,t)$.  In the quantum theory, such a naive treatment sometimes leads to errors \cite{rebhan} because it does not necessarily respect the regularization, and this error does not always vanish when the regulator is taken to infinity.

The key step in linearized perturbation theory is that we solve this problem by working in a different frame, called the kink frame, for all operators and states.  These two frames are related by the unitary operator
\beq
\df={\rm{exp}}\left(-i\int dx f(x)\pi(x)\right) \label{df}
\eeq
which, as desired, transforms the quantum field $\phi(x)$ by
\beq
\df^\dag \phi(x) \df = \phi(x)-f(x).  \label{dfd}
\eeq
Furthermore, it commutes with normal ordering, greatly simplifying calculations.   We transform the Hamiltonian to the kink frame after it is regularized, and so compatibility is assured.

The operator $\df$ maps any zero-kink state to a one-kink state.  That is its usual interpretation as an active transformation.  However, we will instead use it as a passive transformation, to define the kink frame of the Hilbert space, as follows.  

We will refer to the usual identification of the Hilbert space elements with the states as the defining frame.  To make this explicit, we introduce the notation $F$ for a function which takes an element $|\psi\rangle$ in the Hilbert space and yields the physical state $F(|\psi\rangle)$ corresponding to its ray.   Every quantum theory comes with its definition of the function $F$ which identifies rays with states.  Then the kink frame is defined by a second function $F_K$ which provides a different identification of the Hilbert space with the physical states defined by
\beq
F_K(|\psi\rangle)=F(\df|\psi\rangle).
\eeq

Recall that the Hamiltonian $H$ acts on these by generating time evolution, while the momentum
\beq
P=-\int dx \pi(x)\partial_x \phi(x) \label{p}
\eeq
acts on them by generating spatial translations.  
As usual, passive transformations also act on the operators, so that in the kink frame $H\p$ and $P\p$ generate temporal and spatial translations in the kink frame, where
\beq
H\p=\df^\dag H\df\hsp
P\p=\df^\dag P\df=P+\sqrt{Q_0}\pi_0. \label{hp}
\eeq
In particular, as $\df$ is unitary, in the kink frame, the time evolution operator is $e^{-iH\p t}$.  The form of $P\p$ in (\ref{hp}) is easy to interpret.  The $P$ is the momentum stored in all of the mesons, while $\sqrt{Q_0}\pi_0$ term is the kink momentum operator.   This is consistent with the observation above that $\phi_0/\sQ$ is, at linear order, the kink position operator.


What have we gained?  Now the Hamiltonian eigenstates in the one-kink sector are $\df|\psi\rangle$ where $|\psi\rangle$ is defined to be a solution to the eigenvalue problem
\beq
H\p|\psi\rangle=E|\psi\rangle.  \label{se}
\eeq
To see this, from Eqs.~(\ref{hp}) and (\ref{se}) one easily derives
\beq
H\df|\psi\rangle=E\df|\psi\rangle. \label{seo}
\eeq
The equation (\ref{se}), unlike the original equation (\ref{seo}), can be solved in ordinary perturbation theory, despite the fact that the state $\df|\psi\rangle$ is in the one-kink sector.  Thus our procedure for finding Hamiltonian eigenstates  is as follows.  First one solves the equation (\ref{se}) in perturbation theory.  Then one acts on the answer with the nonperturbative operator $\df$ which adds a kink, and finally one identifies this vector with a state using the defining frame $F$.

\subsection{Perturbation Theory}

Now, everything will be expanded in powers of the coupling $g$.  For example, the kink Hamiltonian is expanded
\beq
H\p=\sum_{i=0}^\infty H\p_i \label{semi}
\eeq
where $H\p_i$ consists of terms in the kink Hamiltonian with $i$ powers of the fields when plane-wave normal ordered and a coefficient of order $O(g^{i-2})$.  Similarly the ground state kink energy $Q$ is expanded $Q=\sum_i Q_i$.   

Using Eqs.~(\ref{hp}) and (\ref{cr}) one finds \cite{mekink}
\bea
H\p_0&=&Q_0\hsp H\p_1=0\hsp
H\p_2=Q_1+\frac{\pi_0^2}{2}+\ppin{k} \omega_k B_k^{\ddag} B_k
\nonumber\\
H\p_{n>2}&=&\frac{\sl^{n-2}}{n!}\int dx V^{(n)}(\sl f(x)) : \phi^n(x):_a. \label{hdec}
\eea
As $H\p_0$ is a constant, the perturbation theory begins by considering the eigenstates of $H\p_2$.  These are easily found, as $H\p_2$ is the sum of three terms.  The first, $Q_1$, is a constant.  It is the one-loop correction to the kink mass \cite{cahill76,mekink}.  The second term is the kinetic energy of a particle of mass $Q_0$ and momentum $\sqrt{Q_0}\pi_0$.  As an operator, it corresponds to the quantum mechanics of a free particle, which is simply the center of mass of the kink.  The last is the kinetic energy of all of the mesons, and is simply a sum of quantum harmonic oscillators, representing the various shape and continuum modes.  

The ground state $\vac_0$ of $H\p_2$ is therefore just the ground state of each of these commuting terms
\beq
\pi_0\vac_0=B_k\vac_0=B_S\vac_0=0. \label{v0}
\eeq
The excited $H\p_2$ eigenstates are generated by $B^\ddag$ operators, which excite the various harmonic oscillators, together with boosts.  We will work in the center of mass frame and so will not need the boosts.  For example, one may define the $n$-meson states
\beq
|k_1\cdots k_n \rangle_0=\Bd1\cdots \Bd n \vac_0. \label{2m}
\eeq

Now that the spectrum of $H\p_2$ is known, an arbitrary eigenvector $|\psi\rangle$ of $H\p$ can be constructed.  One first decomposes the state following our perturbative expansion
\beq
|\psi\rangle=\sum_{i=0}^\infty|\psi\rangle_i  \label{semi}
\eeq
where now each $i$ corresponds to a single power of $g$ and $|\psi\rangle_0$ is the corresponding eigenvector of $H\p_2$.  At this point, one might attempt to use standard perturbation theory to solve the eigenvalue equation (\ref{se}).  However, one encounters the standard infrared problem arising from the continuous spectrum that in turn is a consequence of the zero mode.  There are many solutions to this problem, such as promoting the zero mode to a collective coordinate \cite{gjscc}.   This approach is rather cumbersome, as it introduces an infinite number of terms to the Hamiltonian already in the classical theory, which needs to be augmented by another infinite number in the quantum theory \cite{gj76}.

Linearized soliton perturbation theory instead uses a simpler approach.  Imagine that we know $|\psi\rangle_i$ up to some value of $i$.  First, we decompose the state at each order into sectors with different numbers of zero modes
\beq
|\psi\rangle_i=\sum_{n=0}^\infty|\psi\rangle^n_i
\eeq
where $|\psi\rangle^n_i$ contains $\phi_0^n$ when normal-mode normal ordered.  We refer to $|\psi\rangle_i^0$ as the primary component and the rest of the sum as the descendants.  

Next, we impose
\beq
P\p|\psi\rangle=0.
\eeq
Using (\ref{hp}) and the fact that $Q_0$ is of order $O(g^{-2})$, this reduces to the recursion relation
\beq
\pi_0|\psi\rangle_{i+1}=-\frac{1}{\sQ}P|\psi\rangle_i \label{rr}
\eeq
which fixes $|\psi\rangle_{i+1}$ up to terms in the kernel of $\pi_0$.  This kernel consists of the primaries, and so translation-invariance (\ref{rr}) fixes the descendant part of $|\psi\rangle_{i+1}$.  Once this is fixed, then the primary part $|\psi\rangle_{i+1}^0$ can be found using (\ref{se}), as in usual perturbation theory.

\subsection{Leading Correction to the Kink Ground State}

Let us now review the construction of the leading correction $\vac_1$ to the kink ground state $\vac$, following the steps in the previous subsection.  Expanding Eq.~(\ref{p}) using the decomposition (\ref{dec}) one finds
\bea
P
&=&\ppin{k}\Delta_{kB}\left[i\phi_0 \left(-\omega_kB_k^\ddag+\frac{B_{-k}}{2}\right)+\pi_0\left(B_k^\ddag+\frac{B_{-k}}{2\omega_k}\right)\right]\\
&&+i\ppink{2}\Delta_{k_1k_2}\left(-\omega_{k_1}B_{k_1}^\ddag B_{k_2}^\ddag+\frac{B_{-k_1}B_{-k_2}}{4\omega_{k_2}}-\frac{1}{2}\left(1+\frac{\omega_{k_1}}{\omega_{k_2}}\right)B^\ddag_{k_1}B_{-k_2}
\right)\nonumber
\eea
where we have defined the shorthand antisymmetric symbol
\beq
\Delta_{ij}=\int dx \g_i(x)\partial_x \g_j(x).
\eeq
Using Eq.~(\ref{v0}), the right hand ride of the recursion relation (\ref{rr}) is
\beq
P\vac_0=-i\phi_0\ppin{k}\Delta_{kB}\ok{} |k\rangle_0+\frac{i}{2}\ppink{2}\Delta_{k_1k_2}(\ok 2-\ok 1)|k_1k_2\rangle_0.
\eeq
The recursion relation (\ref{rr}) then yields the descendant terms in $\vac_1$
\beq
\vac_1=\vac_1^0-\frac{\phi_0^2}{2\sQ}\ppin{k}\Delta_{kB}\ok{} |k\rangle_0+\frac{i\phi_0}{2\sQ}\ppink{2}\Delta_{k_1k_2}(\ok 2-\ok 1)|k_1k_2\rangle_0.
\eeq

The primary part $\vac_1^0$ can be found using the eigenvalue equation
\beq
(H\p_2-Q_1)\vac_1+H\p_3\vac_0=0 \label{se3}
\eeq
and restricting to the primary subspace of the Hilbert space.  Applying Wick's theorem (\ref{wick}) to the decomposition of $H\p$ in Eq.~(\ref{hdec}), one finds the terms in $H\p_3$ that contribute to the primary part
\bea
H\p_3
&=&\frac{g}{6}\int dx \V3:\phi^3(x):_b+\frac{g}{2}\int dx \V3\I(x)\phi(x)\nonumber\\
&\supset&\frac{g}{6}\ppink{3} V_{k_1k_2k_3}\Bd 1\Bd 2\Bd 3+\frac{g}{2}\ppin{k} V_{\I k}\Bd{}\nonumber
\eea
where we have defined the shorthand symmetric symbol
\beq
V_{\I\stackrel{j}{\cdots}\I,i_1\cdots i_n}=\int \V{n+2j} dx \g_{i_1}(x)\cdots\g_{i_n}(x)\I^j(x).
\eeq
Acting this on $\vac_0$ yields
\beq
H\p_3\vac_0=\frac{g}{6}\ppink{3} V_{k_1k_2k_3}|k_1k_2k_3\rangle_0+\frac{g}{2}\ppin{k} V_{\I k}|k\rangle_0.
\eeq
By Eq.~(\ref{se3}) this must cancel
\bea
(H\p_2-Q_1)\vac_1&=&\frac{\pi_0^2}{2}\vac_1^2+\ppin{k}\ok{} \Bd {} B_k \vac_1^0\\
&=&\frac{1}{2\sQ}\ppin{k}\Delta_{kB}\ok{} |k\rangle_0+\ppin{k}\ok{} \Bd {} B_k \vac_1^0.
\nonumber
\eea
Inverting the operator $\dint\ok{} \Bd {} B_k$, one obtains
\beq
\vac_1^0=-\frac{g}{6}\ppink{3} \frac{V_{k_1k_2k_3}}{\ok 1+\ok 2+\ok 3}|k_1k_2k_3\rangle_0-\frac{g}{2}\ppin{k} \left(\frac{V_{\I k}}{\ok{}}+\frac{\Delta_{kB}}{g\sQ}\right)|k\rangle_0.
\eeq
Note that the operator cannot be inverted on the zero-meson part of $\vac_1^0$, which is proportional to $\vac_0$.  This is just the freedom to normalize the state $\vac$.  We fix this freedom by setting to zero all higher order corrections proportional to $\vac_0$.

\section{Construction of an Asymptotic State} \label{princsez}

\subsection{The Wave Packet}

In this section we will construct an asymptotic state $|\Psi_{x_0}(t=0)\rangle$ corresponding to a localized meson wave packet a distance $|x_0|$ to the left of a kink.  More precisely, if $x=0$ is the location of the center of the kink, then our meson wave packet will correspond to the wave function
\begin{equation}
\Phi(x)=\operatorname{Exp}\left[-\frac{\left(x-x_0\right)^2}{4 \sigma^2}+i x k_0\right], \quad x_0 \ll-\frac{1}{ m}, \quad  \frac{1}{k_0},\frac{1}{m}\ll\sigma \ll\left|x_0\right| .
\end{equation}
Here $k_0$ is the peak momentum.  We will often be interested in the simultaneous limits $m\sigma\rightarrow\infty$ and $m|x_0|\rightarrow\infty$ with $\sigma/x_0\rightarrow 0$, in which the wave packet becomes monochromatic with momentum $k_0$.

We will need to transform this wave packet with respect to the normal mode basis
\begin{equation}
\alpha_\kt=\int d x \Phi(x) \g_\kt(x).  \label{ak}
\end{equation}
In the above limit, the discrete modes vanish and the continuum modes tend to
\bea
\g_k(x)&=&\left\{\begin{tabular}{lll}
$\mb_ke^{-ikx}$&\rm{if} & $x\ll  -1/m$\\
$\md_ke^{-ikx}$&\rm{if} & $x\gg 1/m$\\
\end{tabular}
\right. \label{gk}\\
|\mb_k|^2&=&|\md_k|^2=1\hsp
\mb^*_k=\mb_{-k}\hsp
\md^*_k=\md_{-k}.\nonumber
\eea
In this case one easily evaluates (\ref{ak})
\beq \label{ak}
\alpha_{\kt}=2\sigma\sqrt{\pi}\mb_{\kt}e^{-\sigma^2\left(\kt-k_0\right)^2}e^{i(k_0-\kt)x_0}.
\eeq

Our asymptotic state is defined to be
\beq
|\Phi_{x_0}(t=0)\rangle=\pin{\kt} \alpha_\kt |\kt\rangle^\L. \label{wp}
\eeq
Here $|\kt\rangle^\L$ is the monochromatic state, which we will now construct.  While we will present a particular construction, we note that there are different choices of construction that would lead to the same asymptotic state as the difference between the monochromatic states is annihilated by folding into the wave packet (\ref{wp}).

\subsection{The Monochromatic State}

Recall that we demand that our asymptotic states are translation invariant
\beq
P\p|\Phi_{x_0}(t=0)\rangle=0.
\eeq
We achieve this by demanding the stronger condition
\beq
P\p|\kt\rangle^\L=0.
\eeq
As reviewed above, this condition fixes all of the descendants in $|\kt\rangle^\L$.  However, the primary terms $|\kt\rangle^{\L 0}_i$ are not constrained.

We fix these as follows.  First, at leading order, the monochromatic state should be the bare state defined in Eq.~(\ref{2m})
\beq
|\kt\rangle^{\L}_0=|\kt\rangle_0. \label{kinvar}
\eeq

Each correction $|\kt\rangle_i$ consists of a kink dressed with a cloud of virtual mesons and a meson wave packet dressed with another cloud of virtual mesons.  The key to our construction is that, when $mx_0$ is large, each virtual meson can be associated with either the dressed kink or with the dressed meson wave packet.  We will write states to make this distinction manifest.  In particular, an $n$-meson state
\beq
|k_1\cdots k_j;k_{j+1}\cdots k_n\rangle_0=\Bd{1}\cdots\Bd{n}\vac_0
\eeq
consists of $j$ mesons localized about the wave packet and $n-j$ localized about the kink.  

Roughly, we wish to demand that $|\kt\rangle^\L$ is an eigenvalue of $H\p$.  However, instead of $H\p$ acting on the wave packet, we want to act on it with the left vacuum Hamiltonian $H^\L$ defined by
\beq
H^{\L}_{n\leq 2}=H\p_n\hsp
H^{\L}_{n>2}=\frac{g^{n-2}V^{(n)}_-}{n!}\int dx :\phi^{\L n}(x):_a. \label{vh}
\eeq
Here the left vacuum field is defined by
\beq
\phi^\L(x)=\pin{k}\g^\L_k(x)\left(\Bd{}+\frac{B_{-k}}{2\ok{}}\right)
\eeq
where $\g^\L(x)$ is the asymptotic form of the normal mode on the far left, which in the case of a reflectionless kink is given by the first line in Eq.~(\ref{gk}).  One may wonder whether it is really necessary to replace the ordinary field with the somewhat awkward left vacuum field.  To answer this question, below, we will systematically {\it{not}} replace the $\phi(x)$ in $H\p$ with the $\phi^\L$ in $H^\L$, as it is easily added later.  We will see that the difference between the two is often removed by folding into a wave packet, but that this fails at certain poles corresponding to on-shell processes.   There we will see just what role is played by the vacuum field.  We stress that as a result the intermediate steps below are technically incorrect and should be fixed by replacing $\g$ with $\g^\L$ in $H^\L$.  However, the incorrect part vanishes when folded into the wave packet nearly everywhere, except for a few cases corresponding to contributions of degenerate eigenstates of the kink Hamiltonian which we will discuss when we get to them.  In particular, our final result, Eq.~(\ref{init}), will be correct.


In other words, one uses the kink Hamiltonian but replaces $\V{n}$ with the asymptotic value $V^{(n)}_-$, so that the meson wave packet does not know about the kink.  This is the guiding principle behind our construction.  An incoming meson wave packet must somehow be the same as a wave packet in the absence of a kink.

Concretely the term $H\p|\psi\rangle$ in Eq.~(\ref{se}) is replaced by the following action on each basis vector $|k_1\cdots k_j;k_{j+1}\cdots k_n\rangle_0$ in $|\kt\rangle^\L$
\beq
[H^\L,\Bd{1}\cdots\Bd{j}]\Bd{j+1}\cdots\Bd{n}\vac_0+\Bd{1}\cdots\Bd{j}H\p\Bd{j+1}\cdots\Bd{n}\vac_0
\eeq
so that if 
\beq
|\kt\rangle^\L=\sum \beta_{k_1\cdots k_j;k_{j+1}\cdots k_n}|k_1\cdots k_j;k_{j+1}\cdots k_n\rangle_0
\eeq
then our master definition reads
\beq
E|\kt\rangle^\L=\sum \beta_{k_1\cdots k_j;k_{j+1}\cdots k_n}\left(
[H^\L,\Bd{1}\cdots\Bd{j}]+\Bd{1}\cdots\Bd{j}H\p
\right)\Bd{j+1}\cdots\Bd{n}\vac_0. \label{princ}
\eeq
It is understood that the right and left hand sides are restricted to the primary states, so that it defines $|\kt\rangle^{\L 0}$.  Were it imposed on the descendants, it would violate (\ref{kinvar}), although translation-invariance might be restored by folding into the wave packet.

\subsection{Consistency}

The variable $E$ is not assumed to be an eigenvalue of any operator.  We will see that it is the sum of the energy of the ground state kink plus the vacuum energy of a moving meson that would be evaluated in the vacuum sector.  The equation (\ref{princ}) defines $|\kt\rangle^{\L 0}_{i+1}$ given $|\kt\rangle_i$.  However, one also needs to assign each meson to either the meson wave packet or the kink.  Clearly if the meson at order $i+1$ arose from $H^\L$ acting on wave packet mesons, or $H\p$ acting on the mesons in the kink's cloud at order $i$, then the order $i+1$ meson should be assigned to the wave packet or the kink cloud respectively.  However, it may be that it arises from the commutator of $H^\L$ with the wave packet creation operators which then is contracted with a meson in the dressed kink.  It is an important check of the consistency of this prescription that such mesons vanish when folded into the wave packet, and so do not contribute to our asymptotic state.  Below we will see that this is indeed the case.

Also, we need to check that this definition has the properties stated in the introduction.  In particular, we want the asymptotic state to evolve via a rigid translation of the entire meson wave packet with respect to the kink.  This is nontrivial, as the wave packet is constructed via a hybrid of $H^\L$ and $H\p$ whereas temporal evolution is generated by $H\p$ alone.  Again, it is necessary that the mismatch between the two Hamiltonians be annihilated when $|\kt\rangle^\L$ is folded into the wave packet.  We will also see that this is the case, and the above prescription indeed leads to a wave packet which moves rigidly when evolved with respect to $H\p$.


\section{Leading Correction to the One-Meson Asymptotic State} \label{gsez}

In this section we will construct the leading quantum correction $|\kt\rangle^\L_1$ to the monochromatic part $|\kt\rangle^\L$ of an asymptotic state consisting of a kink and a nearly-monochromatic meson wave packet far to its left.  Recall that the state is expanded in powers of $g$ with the first power being the corresponding eigenstate $|\kt\rangle_0$ of $H\p_2$
\beq
|\kt\rangle^\L=\sum_{i=0}^\infty |\kt\rangle_i^\L\hsp |\kt\rangle^\L_0=|\kt\rangle_0.
\eeq

\subsection{The Descendants from Translation Invariance}

Following our construction in Sec.~\ref{princsez}, the descendant part of the state, the part with powers of $\phi_0$, is fixed by demanding translation invariance 
\beq
P\p|\kt\rangle^\L=0.
\eeq

This is equivalent to the recursion relation
\beq
P|\kt\rangle_i^\L=-\sQ\pi_0|\kt\rangle_{i+1}^\L
\eeq
whose left hand side is
\bea
P|\kt\rangle_0^\L=P|\kt\rangle_0&=&-i\phi_0\ppin{k}\Delta_{kB}\ok{} |\kt;k\rangle_0+\frac{i}{2}\ppink{2}\Delta_{k_1k_2}(\ok 2-\ok 1)|\kt;k_1k_2\rangle_0\nonumber\\
&&+\frac{i\phi_0}{2}\Delta_{-\kt,B}\vac_0+\frac{i}{2}\ppin{k}\Delta_{-\kt,k}\left(1+\frac{\ok{}}{\okt{}}\right)|k\rangle_0.
\eea

Inverting the $\pi_0$ we find the descendants
\bea
|\kt\rangle^\L_1&=&|\kt\rangle^{\L0}_1-\frac{\phi^2_0}{2\sQ}\ppin{k}\Delta_{kB}\ok{} |\kt;k\rangle_0+\frac{\phi_0}{2\sQ}\ppink{2}\Delta_{k_1k_2}(\ok 2-\ok 1)|\kt;k_1k_2\rangle_0\nonumber\\
&&+\frac{\phi^2_0}{4\sQ}\Delta_{-\kt,B}\vac_0+\frac{\phi_0}{2\sQ}\ppin{k}\Delta_{-\kt,k}\left(1+\frac{\ok{}}{\okt{}}\right)|k\rangle_0. \label{kts}
\eea

\subsection{The Primaries}

Next we find the primary part $|\kt\rangle_1^{\L 0}$, using our master formula (\ref{princ}) which at leading order is just
\beq
(H\p_{2}-Q_1-\okt{})|\kt\rangle^\L_1+\left[H^{\L}_{3},B^\dag_\kt\right]\vac_0+B^\dag_\kt H\p_3\vac_0=0 \label{h3eq}
\eeq
where the vacuum Hamiltonian is
\bea
H^\L_3&=&\frac{g}{6}\int dx V^{(3)}_-:\phi^3(x):_a
=\frac{g V^{(3)}_-}{6}\int dx \left[:\phi^3(x):_b+3\I(x)\phi(x)\right]. 
\eea
We remind the reader that the vacuum Hamiltonian, as defined in Eq.~(\ref{vh}), should contain the vacuum field $\phi^\L(x)$ and not $\phi(x)$, but that we are intentionally using the wrong form in our derivation to see where the vacuum field will be necessary. 

The third term in (\ref{h3eq}) is then
\beq
B^\dag_\kt H\p_3\vac_0\Big|_{m=0}=\frac{g}{6}\ppink{3} V_{k_1k_2k_3}|\kt;k_1k_2k_3\rangle_0+\frac{g}{2}\ppin{k} V_{\I k}|\kt;k\rangle_0.
\eeq
The only term in the vacuum Hamiltonian which will contribute to $|\kt\rangle_1^{\L 0}$ is
\beq
H^\L_3\supset  \frac{g V^{(3)}_-}{4}\int dx \left[
\ppink{3} \frac{\g_{k_1}(x) \g_{k_2}(x) \g_{k_3}(x)}{\ok 3} \Bd 1 \Bd 2 B_{-k_3}
+\I(x)\ppin{k} \frac{\g_k(x)}{\ok {}} B_{-k}
\right].
\eeq
As a result the second term in (\ref{h3eq}) is 
\beq
\left[H^{\L}_{3},B^\dag_\kt\right]\vac_0=\frac{g V^{(3)}_-}{4\okt{}}\int dx \left[
\I(x){\g_{-\kt}(x)}{} \vac_0+\ppink{2} {\g_{k_1}(x) \g_{k_2}(x) \g_{-\kt}(x)} |k_1k_2\rangle_0
\right].
\eeq
Finally, the first term is
\bea
(H\p_{2}-Q_1-\okt{})|\kt\rangle^\L_1&=&\left(\ppin{k}\ok{}B^\dag_k B_k-\okt{}\right) |\kt\rangle^{\L0}_1+\frac{\pi_0^2}{2} |\kt\rangle^{\L2}_1\\
&&\hspace{-4cm}=\left(\ppin{k}\ok{}B^\dag_k B_k-\okt{}\right) |\kt\rangle^{\L0}_1+\frac{1}{2\sQ}\ppin{k}\Delta_{kB}\ok{} |\kt;k\rangle_0-\frac{1}{4\sQ}\Delta_{-\kt,B}\vac_0.\nonumber
\eea
Combining these terms, Eq.~(\ref{h3eq}) reduces to
\bea
\left(\ppin{k}\ok{}B^\dag_k B_k-\okt{}\right)|\kt\rangle^{\L0}_1&=&\left[\frac{\Delta_{-\kt,B}}{4\sQ}
-\frac{g V^{(3)}_-}{4\okt{}}\int dx\I(x){\g_{-\kt}(x)}{} 
\right]\vac_0\\
&&\hspace{-5cm}-\frac{g}{6}\ppink{3} V_{k_1k_2k_3}|\kt;k_1k_2k_3\rangle_0-\frac{g}{2}\ppin{k} \left[V_{\I k}+\frac{\Delta_{kB}\ok{} }{g\sQ}
\right]|\kt;k\rangle_0\nonumber\\
&&\hspace{-5cm}-\frac{g V^{(3)}_-}{4\okt{}}\int dx 
\ppink{2} {\g_{k_1}(x) \g_{k_2}(x) \g_{-\kt}(x)} |k_1k_2\rangle_0.
\nonumber
\eea
The kernel of the operator on the left consists of states with an on-shell energy of $\okt{}$.  Indeed, our definition of the state is ambiguous in that one may add $|\kt\rangle_0$, reflecting the freedom to change the normalization, and also degenerate eigenstates such as $|-\kt\rangle$ or those with multiple on-shell mesons.  These on-shell meson states will travel at different velocities, and so we do not include them.   This choice of inverse arises at each order and should be considered an integral part of our construction of the monochromatic state.

We therefore choose the following inverse
\bea
|\kt\rangle^{\L0}_1&=&\left[-\frac{\Delta_{-\kt,B}}{4\sQ\okt{}}
+\frac{g V^{(3)}_-}{4\okt{}^2}\int dx\I(x){\g_{-\kt}(x)}{} 
\right]\vac_0-\frac{g}{2}\ppin{k} {\left[\frac{V_{\I k}}{\ok{}}+\frac{\Delta_{kB}}{g\sQ}
\right]}{}|\kt;k\rangle_0\nonumber\\
&&\hspace{-0cm}-\frac{g}{6}\ppink{3} \frac{V_{k_1k_2k_3}}{\ok1+\ok2+\ok3}|\kt;k_1k_2k_3\rangle_0\nonumber\\
&&\hspace{-0cm}+\frac{g V^{(3)}_-}{4\okt{}}\int dx 
\ppink{2} \frac{\g_{k_1}(x) \g_{k_2}(x) \g_{-\kt}(x)}{\okt{}-\ok1-\ok2} |k_1k_2\rangle_0. \label{ktp}
\eea
This completes our computation of the leading order correction $|\kt\rangle^\L_1$ to $|\kt\rangle^\L$.

\subsection{Folding $|\kt\rangle^\L$ Into a Wave Packet}

To obtain the asymptotic state, we must fold $|\kt\rangle^\L$ into the wave packet (\ref{wp}).  We will further decompose all states $|\psi\rangle$ as
\beq
|\psi\rangle_i^m=\sum_{n=0}^\infty |\psi\rangle_i^{mn}
\eeq
where $|\psi\rangle_i^{mn}$ is the part of the $i$th order term proportional to $\phi_0^m$ with $n$ mesons, or more precisely, with $n$ $B^\ddag$ operators acting on $\vac_0$.    We will now decompose our monochromatic asymptotic state in Eqs.~ (\ref{kts}) and (\ref{ktp}) into sectors with various numbers of mesons, and fold them into (\ref{wp}) one at a time.

\subsubsection{The Zero-Meson Sector}

The zero-meson part of Eq.~(\ref{kts}) with two zero modes is 
\beq
|\kt\rangle^{\L 20}_1=\frac{\phi^2_0}{4\sQ}\Delta_{-\kt,B}\vac_0.
\eeq
Folding it into the wave packet (\ref{wp}) one arrives at
\bea
\pin{\kt} \alpha_\kt |\kt\rangle^{\L 20}_1 
&=&\pin{\kt} 2\sigma\sqrt{\pi}\mb_{\kt}e^{-\sigma^2\left(\kt-k_0\right)^2}e^{i(k_0-\kt)x_0} \frac{\phi^2_0}{4\sQ}\Delta_{-\kt,B}\vac_0\\
&=& \frac{2\sigma\sqrt{\pi}\phi^2_0}{4\sQ} e^{ik_0x_0} \vac_0\int dx \g\p_B(x) \pin{\kt} \g_{-\kt}(x) \mb_{\kt}e^{-\sigma^2\left(\kt-k_0\right)^2}e^{-i\kt x_0}.\nonumber
\eea
As $|x_0|$ is much larger than both $\sigma$ and also $1/m$, the only length scale which appears in $\g_{-k}(x)$, the argument of the $e^{-i\kt x_0}$ changes more quickly in $\kt$ than any other term unless $x\sim x_0$, in which case the argument of $\g_{-k}(x)$ changes at the opposite rate.  Thus, the $\kt$ integral is exponentially suppressed in $(x-x_0)$.  As a result, all but an exponentially small portion of this quantity arises from $x\sim x_0$, where $\g_B(x)$ is itself exponentially suppressed in $mx_0$.  Thus, in the $mx_0\rightarrow\infty$ limit, we conclude that $|\kt\rangle^{\L 20}_1$ vanishes when folded into the wave packet (\ref{wp}).

The zero-meson part of Eq.~(\ref{kts}) with no zero modes is 
\beq
|\kt\rangle^{\L00}_1=\left[-\frac{\Delta_{-\kt,B}}{4\sQ\okt{}}
+\frac{g V^{(3)}_-}{2\okt{}^2}\int dx\I(x){\g_{-\kt}(x)}{} 
\right]\vac_0.
\eeq
As $\okt{}$ is essentially constant in the support of the Gaussian $e^{-\sigma^2\left(\kt-k_0\right)^2}$, it may be replaced with $\ok{0}$ and be pulled out of the $\kt$ integral.  Then, the same argument as above shows that the first term in $|\kt\rangle^{\L00}_1$ does not contribute to the wave packet.  

What about the second term?
\bea
\pin{\kt} \alpha_\kt \left[\frac{g V^{(3)}_-}{2\okt{}^2}\int dx\I(x){\g_{-\kt}(x)}{} 
\vac_0\right]&&\\
&&\hspace{-5cm}={\sigma\sqrt{\pi}g V^{(3)}_-}{}\int dx\I(x)\pin{\kt}\frac{\g_{-\kt}(x)}{\okt{}^2}\mb_{\kt}e^{-\sigma^2\left(\kt-k_0\right)^2}e^{i(k_0-\kt)x_0}\nonumber\\
&&\hspace{-5cm}=\frac{\sigma\sqrt{\pi}g V^{(3)}_- \mb_{k_0}e^{ik_0x_0}}{\ok{0}^2}\int dx\I(x)\pin{\kt}{\g_{-\kt}(x)}{}e^{-\sigma^2\left(\kt-k_0\right)^2}e^{-i\kt x_0}.\nonumber
\eea
Again the $\kt$ integral has support at $x\sim x_0$ where $\I(x)$ is exponentially suppressed in $mx_0$.  As a result, this last contribution to the zero-meson sector also vanishes when folded into the wave packet.

\subsubsection{The One-Meson Sector}

There is only a single term in $|\kt\rangle^\L_1$ in the one-meson sector
\beq
|\kt\rangle^{\L 11}_1=\frac{\phi_0}{2\sQ}\ppin{k}\Delta_{-\kt,k}\left(1+\frac{\ok{}}{\okt{}}\right)|k\rangle_0.
\eeq
Folding this into the wave packet one obtains
\bea
\pin{\kt} \alpha_\kt |\kt\rangle^{\L 11}_1
&=&
\frac{\sigma\sqrt{\pi} \phi_0}{\sQ}
\pin{\kt}
\mb_{\kt}e^{-\sigma^2\left(\kt-k_0\right)^2}e^{i(k_0-\kt)x_0}
\ppin{k}\Delta_{-\kt,k}\left(1+\frac{\ok{}}{\okt{}}\right)|k\rangle_0\nonumber\\
&&\hspace{-3cm}=
\frac{\sigma\sqrt{\pi} \phi_0}{\sQ}
\int dx 
\ppin{k} \g^\prime_k(x)
\pin{\kt}  \g_{-\kt}(x)
\mb_{\kt}e^{-\sigma^2\left(\kt-k_0\right)^2}e^{i(k_0-\kt)x_0}
\left(1+\frac{\ok{}}{\okt{}}\right)|k\rangle_0
\eea
Again the $\kt$ integral is only supported at $x\sim x_0$ where we can approximate the normal modes $\g$ by plane waves (\ref{gk}), leading to
\bea
\pin{\kt} \alpha_\kt |\kt\rangle^{\L 11}_1
&=&
-\frac{i\sigma\sqrt{\pi} \phi_0}{\sQ}
\pin{k} k\mb_k \label{i1}\\
&&\times
\pin{\kt}    
e^{-\sigma^2\left(\kt-k_0\right)^2}e^{i(k_0-\kt)x_0}
\left(1+\frac{\ok{}}{\okt{}}\right)\left[\int dx e^{i(\kt-k) x}\right]|k\rangle_0\nonumber\\
&=&
-\frac{2i\sigma\sqrt{\pi} \phi_0}{\sQ}
\pin{\kt} \kt\mb_\kt 
e^{-\sigma^2\left(\kt-k_0\right)^2}e^{i(k_0-\kt)x_0}
|\kt\rangle_0\nonumber\\
&=&\pin{\kt} 
\alpha_\kt |\kt\rangle^{\L P}_1
\hsp
|\kt\rangle^{\L P}_1=
-\frac{i\kt \phi_0}{\sQ}|\kt\rangle_0.
\nonumber
\eea
In other words, when folded into the wave packet, $|\kt\rangle^{\L 11}_1$ and $|\kt\rangle^{\L P}_1$ are equal.  However, we will now see that the later has a simple, physical interpretation.

Consider the expectation value of $\phi(x)$ in the leading order vacuum frame kink ground state $\df\vac_0$
\bea
{}_{\ \ 0}^{}\langle 0|\df^\dag \phi(x)\df\vac_0^{}&=&{}_{\ \ 0}^{}\langle 0|(f(x)+\phi(x))\vac_0^{}\nonumber\\
&=&{}_{\ \ 0}^{}\langle 0|(f(x)+\phi_0\g_B(x))\vac_0^{}={}_{\ \ 0}^{}\langle 0|\left(f(x)-\frac{\phi_0}{\sQ}f\p(x)\right)\vac_0^{}\nonumber\\
&=&{}_{\ \ 0}^{}\langle 0|\left(f\left(x-\frac{\phi_0}{\sQ}\right)+O\left(\frac{1}{Q_0}\right)\right)\vac_0^{}. \label{pos}
\eea
We thus see that the kink profile, to leading order, is $f(x-\phi_0/\sQ)$, and so to leading order $\phi_0/\sQ$ is the position of the kink.  More precisely, it is the eigenvalue of $\phi_0$ divided by $\sQ$ which provides the kink position in a $\phi_0$ eigenstate.

Eq.~(\ref{i1}) tells us that, at leading order, the wave packet is corrected by $(-i\kt)$ times the position of the kink.  Of course, $(-i\kt)$ is just the spatial derivative of the wave packet itself, and so $|\kt\rangle^{\L P}_1$ is the first order Taylor series expansion of a translation of the meson wave packet.  Thus we conclude that, as a result of this term, when the kink is moved, the meson is moved by just the same amount.

Recall that the term contains powers of $\phi_0$ and so is a descendant, and the descendants were derived using translation-invariance of the combined meson and kink system.  Thus translation-invariance has implied that the relative distance between the kink and the meson wave packet is fixed.  More precisely, the kink is in a momentum eigenstate which corresponds to an infinite superposition of position eigenstates, with every possible position summed over.  We learn that in each of these position eigenstates, the kink-meson distance is fixed.  Of course our perturbative treatment breaks down for positions too far from zero, and so our expressions are perturbative in $\phi_0/\sQ$.

\subsubsection{The Two-Meson Cloud}

The last interesting term in $|\kt\rangle^\L_1$ is
\beq
|\kt\rangle^{\L 02}_1\supset
-\frac{g V^{(3)}_-}{4\okt{}}\int dx 
\ppink{2} \frac{\g_{k_1}(x) \g_{k_2}(x) \g_{-\kt}(x)}{\okt{}-\ok1-\ok2} |k_1k_2\rangle_0.
\eeq
Folding this into the wave packet, the same arguments as above imply that the $\kt$ integral is supported at $x\sim x_0$ where the normal modes are plane waves, and so we find that 
\bea
\pin{\kt} \alpha_\kt |\kt\rangle^{\L 02}_1 &\supset&
\pin{\kt} \alpha_\kt \left[ -\frac{g V^{(3)}_-}{4\okt{}}\int dx 
\ppink{2} \frac{\g_{k_1}(x) \g_{k_2}(x) \g_{-\kt}(x)}{\okt{}-\ok1-\ok2} |k_1k_2\rangle_0\right]\label{split}\\
&&\hspace{-2cm}=\pin{\kt} \alpha_\kt \left[ -\frac{g V^{(3)}_-}{4\okt{}}
\ppink{2} \frac{\mb_{k_1} \mb_{k_2}\mb_{-\kt}2\pi\delta(\kt-k_1-k_2)}{\okt{}-\ok1-\ok2} |k_1k_2\rangle_0\right]\nonumber\\
&&\hspace{-2cm}=\pin{\kt} \alpha_\kt |\kt\rangle^{\L S}\hsp
|\kt\rangle^{\L S}= -\frac{g V^{(3)}_-}{4\okt{}}
\ppin{k} \frac{\mb_{k} \mb_{\kt-k} \mb_{-\kt}}{\okt{}-\ok{}-\omega_{\kt-k}} |k,\kt-k\rangle_0.\nonumber
\eea
We conclude that, after being folded into the wave packet, this term in $|\kt\rangle^{\L 02}_1$ is equal to $|\kt\rangle^{\L S}$.

This argument fails at the pole $\okt{}=\ok1+\ok2$.  Here one cannot assume that the denominator varies slowly as compared with the exponential at large $x_0$, as $x_0$ would need to be larger than the inverse distance to the pole which is unbounded.  This is the first place where the vacuum field $\phi^\L(x)$ in the definition (\ref{vh}) of $H^\L$ is necessary.  Recalling that $H^\L$ contains $\phi^\L(x)$ and not $\phi(x)$, the three factors of $\g$ in the first line of (\ref{split}) are replaced directly with plane waves, leading to the second line with no need for an argument involving $\alpha_k$.  Physically this is very important.  Had $H^\L$ contained $\phi(x)$ and not $\phi^\L(x)$ then the pole would have led to a finite remainder consisting of two on-shell mesons.  This choice of initial state therefore contains a physical mixture of 2-meson states, and so the probability of meson multiplication would be nonzero even before the meson approached the kink.  Clearly this is an inappropriate initial condition for calculating the meson multiplication resulting from kink-meson scattering, as the mesons that need to be created are already present in the initial state.  Also, these on-shell mesons will travel more slowly then the rest of the wave packet, and so our first criterion for the definition of an asymptotic state would not be satisfied had we used $\phi(x)$ and not $\phi^\L(x)$ in the definition (\ref{vh}) of the vacuum Hamiltonian.

This meson-splitting term also has a simple, physical explanation.  Far from the kink the meson may split into two mesons if $V^{(3)}_-$ is nonzero.  However, far from the kink, the mesons must conserve momentum separately from the kink, and so the total momentum is always $\kt$.  This means that the two-meson component of the meson wave function is far off-shell, and these two mesons are quite virtual.  In this sense, the two-meson cloud shown here is quite similar to the cloud which appears around an isolated meson in the vacuum sector.

We now arrive at the first application of this work.  The coefficients in the term $|\kt\rangle^{\L S}$ are given in the first expression in Eq.~(6.4) of Ref.~\cite{memult}.  In that reference they were found by simply replacing the normal modes with plane waves and removing the terms localized near the kink.  Here we have, instead, derived that result from a choice of initial state.  More importantly, our initial state exactly preserves translation-invariance and so the reduced norms used in the calculation of the meson multiplication probability may be applied.  In contrast, since $H^\L$ does not commute with the momentum operator, an exact eigenstate of $H^\L$, as proposed in Ref.~\cite{memult}, would not be translation invariant.  We have now avoided this problem by adding corrections to the $H^\L$ eigenstate which are exponentially suppressed in $mx_0$.

Note that the second expression in Eq.~(6.4), describing the four-meson component which does not contribute to meson multiplication, does not agree with our asymptotic state (\ref{ktp}).  This is because here we use the full kink Hamiltonian to determine the meson cloud about the kink, unlike the choice in that paper.

\subsubsection{The Multi-Meson Sectors}

The multimeson terms $n\geq 2$ in $|\kt\rangle$, as reported in Eqs.~(\ref{kts}) and (\ref{ktp}), can be summarized as follows, up to subleading order
\beq
|\kt\rangle^\L\Big|_{n\geq 2}=\left(B^\dag_{\kt}-
\frac{g V^{(3)}_-}{4\okt{}}
\pin{k} \frac{\mb_{k} \mb_{\kt-k} \mb_{-\kt}}{\okt{}-\ok{}-\omega_{\kt-k}} B^\dag_kB^\dag_{\kt-k}
\right)\vac.
\eeq
Here the meson splitting term $|\kt\rangle^{\L S}$ is included by dressing the perturbative meson creation operator $B^\dag_\kt$.

Assembling all of our results, we conclude that the wave packet (\ref{wp}) may be written, up to order $O(g)$, as
\bea
|\Phi_{x_0}(t=0)\rangle&=&\pin{\kt} \alpha_\kt 
\left[|\kt\rangle^\L\Big|_{n\geq 2}+|\kt\rangle^{\L P}
\right] \label{init}\\
&&\hspace{-2cm}=\pin{\kt} \alpha_\kt 
\left[\left(B^\dag_{\kt}-
\frac{g V^{(3)}_-}{4\okt{}}
\pin{k} \frac{\mb_{k} \mb_{\kt-k} \mb_{-\kt}}{\okt{}-\ok{}-\omega_{\kt-k}} B^\dag_kB^\dag_{\kt-k}\right)\vac
-\frac{i\kt \phi_0}{\sQ}|\kt\rangle_0
\right].\nonumber
\eea
This simple formula is one of our main results.  The vacuum Hamiltonian only enters via the coefficient $V^{(3)}_-$ which appears in the dressing of the meson creation operator.  This is physically reasonable, the meson is far from the kink and so is not yet affected by the full interactions of the kink Hamiltonian.  However $\vac$ is the perturbative ground state of the full kink Hamiltonian, which again is reasonable as the kink is affected by the kink, and so is its meson cloud.  

In summary, the effects of the vacuum and kink Hamiltonians at this order are clearly separated in the wave packet state (\ref{init}).  The vacuum Hamiltonian determines the dressed meson creation operator, and so appears inside of the round brackets, while the state $\vac$, representing the dressed kink on which this terms acts, is determined by the full kink Hamiltonian $H\p$.  

This agrees with the naive intuition that the meson wave packet behavior is captured by the vacuum Hamiltonian and the kink, together with its meson cloud, is captured by the kink Hamiltonian.  Of course, both evolve via the action of the kink Hamiltonian, as it is the defining Hamiltonian (\ref{ham}), written in the kink frame.  In the next section we will show that this decomposition is in fact preserved by evolution under the full kink Hamiltonian.



\section{Evolving the Asymptotic State} \label{movesez}

After a time $t$, the initial state $|\Phi_{x_0}(t=0)\rangle$ evolves to
\beq
|\Phi_{x_0}(t)\rangle=e^{-iH\p t}|\Phi_{x_0}(t=0)\rangle. \label{schrod}
\eeq
In this section we will evaluate this at leading and subleading orders.

\subsection{Leading Order} \label{leadmove}

At leading order the evolved wave packet is
\beq
|\Phi_{x_0}(t)\rangle_0=e^{-iH\p_2 t}|\Phi_{x_0}(t=0)\rangle_0=\pin{\kt} \alpha_\kt e^{-iH\p_2 t} |\kt\rangle_0.
\eeq
The leading order evolution of each $|\kt\rangle_0$ is
\beq
e^{-iH\p_2 t} |\kt\rangle_0=e^{-i(Q_1+\okt{}) t} |\kt\rangle_0.
\eeq
Such a $\kt$ dependence, following standard arguments which we will now review, yields rigid motion of the wave packet.

Expanding $\omega$ about $k_0$ one finds
\beq
\okt{}=\ok{0}+\frac{\kt_0}{\okt{}}(\kt-k_0)
\eeq
up to corrections of order $O\left((\kt-k_0)^2\right)$ which yield wave packet spreading but vanish at large $\sigma$.  Here $\kt_0/\okt{}$ is the expected velocity of the wave packet, and its position at time $t$ is
\beq
x_t=x_0+\frac{\kt_0}{\okt{}}t.
\eeq
Dropping the higher order corrections, and so ignoring wave packet spreading, one finds
\bea
|\Phi_{x_0}(t)\rangle_0&=&\pin{\kt} 2\sigma\sqrt{\pi}\mb_{\kt}e^{-\sigma^2\left(\kt-k_0\right)^2}e^{i(k_0-\kt)x_0} e^{-i\left(Q_1+\ok{0}+\frac{\kt_0}{\okt{}}(\kt-k_0)\right) t} |\kt\rangle_0\\
&=&2\sigma\sqrt{\pi}e^{-i\left(Q_1+\ok{0}\right) t}\pin{\kt} \mb_{\kt}e^{-\sigma^2\left(\kt-k_0\right)^2}e^{i(k_0-\kt)x_t}  |\kt\rangle_0
\nonumber\\
&=&e^{-i\left(Q_1+\ok{0}\right) t}|\Phi_{x_t}(t=0)\rangle_0
.\nonumber
\eea
This last expression means that, at leading order, at time $t$ the initial meson wave packet is only changed, up to an overall phase and ignoring the usual spreading, by the substitution $x_0\rightarrow x_t$.   Thus, at leading order we conclude that these states satisfy the first property that we require, while the second property was already manifestly satisfied as a result of the choice of descendants.

However at this order we have not included our corrections to $|\kt\rangle^\L$ or to $e^{-iH\p t}$, so this has been rather trivial.  Now we will check that this property is maintained at the next order.

\subsection{Subleading Order: Descendants}

At order $O(g)$, two kinds of terms appear in the evolution equation (\ref{schrod}).  Either the evolution operator $e^{-iH\p t}$ is at order $O(g)$ and the initial state $|\Phi_{x_t}(t=0)\rangle_0$ is at order $O(g^0)$ or else the evolution operator $e^{-iH\p_2 t}$ is at order $O(g^0)$ and the initial state $|\Phi_{x_t}(t=0)\rangle_1$ is at order $O(g^1)$.

In the case of the first terms, the evolution operator may be expanded
\bea
e^{-iH\p t}\Big|_{O(g)}&=&\sum_{n=0}^\infty \frac{(-it)^n}{n!}H^{\prime n}\Big|_{O(g)}=\sum_{n=0}^\infty \frac{(-it)^n}{n!}\sum_{m=0}^{n-1}H^{\prime m}_2  H^{\prime}_3H^{\prime n-m-1}_2. \label{id}
\eea
We will often simplify this using
\bea
\sum_{n=0}^\infty \frac{(-it)^n}{n!}\sum_{m=0}^{n-1} a^m b^{n-1-m}
=\frac{e^{-ibt}-e^{iat}}{b-a}.
\eea

\subsubsection{$\phi_0^3$ Terms}

We will begin with the terms containing three powers of $\phi_0$.  These can only arise from the term 
\beq
H\p_3\supset \frac{g V_{BBB}}{6}\phi_0^3.
\eeq

Using
\beq
\partial_x\V2=\V3 gf\p(x)=- g\sQ\V3 \g_B(x)
\eeq
and the Sturm-Liouville equation for the normal modes
\beq
\V2 \g_B(x)=\partial_x^2 \g_B(x)
\eeq
one finds
\bea
V_{BBB}&=&\int dx \V3 \g^3_B(x)=-\frac{1}{g\sQ} \int dx \partial_x(\V2) \g^2_B(x)\\
&=&\frac{2}{g\sQ} \int dx \V2 \g_B(x) \g^\prime_B(x)=\frac{1}{g\sQ} \int dx \partial_x (\g^2_B(x))=0.\nonumber
\eea
Therefore there is no term in $H\p_3$ of order $\phi_0^3$, and so no contribution to the $O(g)$ evolution with three powers of $\phi_0$.

\subsubsection{$\phi_0^2$ Terms}

The evolution operator $e^{-iH\p t}$ at order $O(g)$ contains terms of order $\phi_0^2$ arising from
\beq
H\p_3\supset \frac{g\phi_0^2}{2} \ppin{k} V_{BBk} \left(\Bd{}+\frac{B_{-k}}{2\ok{}}
\right). \label{h32}
\eeq
Let us first consider the $\Bd{}$ term, plugging it into (\ref{id}) to evaluate
\bea
e^{-iH\p t}\Big|_{O(g)}|\kt\rangle_0&\supset& \frac{g\phi_0^2}{2} \ppin{k} V_{BBk} \sum_{n=0}^\infty \frac{(-it)^n}{n!}\sum_{m=0}^{n-1}(Q_1+\okt{}+\ok{})^m  \Bd{} (Q_1+\omega_{\kt})^{n-m-1}|\kt\rangle_0\nonumber\\
&=& -\frac{g\phi_0^2}{2} \ppin{k} \frac{V_{BBk}}{\ok{}} \sum_{n=0}^\infty \frac{(-it)^n}{n!}(Q_1+\omega_{\kt})^{n} \left[1-\left(\frac{Q_1+\okt{}+\ok{}}{Q_1+\omega_{\kt}}\right)^n\right]{}  |\kt;k\rangle_0\nonumber\\
&=& \frac{g\phi_0^2}{2} e^{-i(Q_1+\omega_{\kt})t} \ppin{k} \frac{V_{BBk}}{\ok{}} \left(e^{-i\ok{}t}- 1\right)|\kt;k\rangle_0.
\eea
To simplify further, using the Sturm-Liouville equation
\beq
\V2 \g_k(x)=\g\pp_k(x)+\ok{}^2\g_k(x)
\eeq
to derive 
\bea
V_{BBk}&=&\int dx \V3 \g^2_B(x) \g_k(x)=\frac{1}{g\sQ}\int dx \V2 (\g_B(x)\g\p_k(x)+\g_k(x)\g\p_B(x))\nonumber\\
&=&\frac{1}{g\sQ}\int dx (\partial_x(\g\p_k(x)\g\p_B(x))+\ok{}^2\g_k(x)\g\p_B(x))=\frac{\ok{}^2\Delta_{kB}}{g\sQ}
\eea
the term above is
\bea
e^{-iH\p t}\Big|_{O(g)}|\kt\rangle_0&\supset&
\frac{\phi_0^2}{2\sQ} e^{-i(Q_1+\omega_{\kt})t} \ppin{k} {\ok{} \Delta_{kB}}{} \left(e^{-i\ok{}t}- 1\right)|\kt;k\rangle_0. \label{c22a}
\eea

As we will see is often the case below, for each such term, there is a corresponding term which is $O(g^0)$ in the evolution operator $e^{-iH\p_2 t}$ and $O(g)$ in $|\kt\rangle^\L_1$.  Using Eq.~(\ref{kts}) one finds the term
\bea
e^{-iH\p_2 t}|\kt\rangle^{\L 22}_1&=&e^{-iH\p_2 t}\left[-\frac{\phi^2_0}{2\sQ}\ppin{k}\Delta_{kB}\ok{} |\kt;k\rangle_0\right]\label{c22b}\\
&=&-e^{-iQ_1 t}\frac{\phi^2_0}{2\sQ}\ppin{k}\Delta_{kB}\ok{} e^{-i(\okt{}+\ok{}) t}|\kt;k\rangle_0.\nonumber
\eea

Adding the contributions in Eqs.~(\ref{c22a}) and (\ref{c22b}) one finds the total 2-meson contribution to the $\phi_0^2$ terms
\beq
e^{-iH\p t}|\kt\rangle \supset -e^{-i(Q_1+\okt{}) t}\frac{\phi^2_0}{2\sQ}\ppin{k}\Delta_{kB}\ok{} |\kt;k\rangle_0=e^{-i(Q_1+\okt{})t}|\kt\rangle^{\L 22}_1.
\eeq
Therefore these terms evolve by a multiplication by the phase $e^{-i(Q_1+\okt{})t}$.  Following the same argument as in the leading order case in Subsec.~\ref{leadmove}, this implies that this term in the wave packet moves rigidly as time passes, up to wave packet spreading corrections.

Let us turn now to the zero-meson terms that are proportional to $\phi_0^2$.  Plugging the second term in Eq.~(\ref{h32}) into Eq.~(\ref{id}) one finds
\bea
e^{-iH\p t}|\kt\rangle_0&\supset&
\frac{g\phi_0^2}{4} \ppin{k} V_{BBk} 
\sum_{n=0}^\infty \frac{(-it)^n}{n!}\sum_{m=0}^{n-1}Q_1^{m}  \frac{B_{-k}}{\ok{}} (Q_1+\okt{})^{n-m-1} |\kt\rangle_0\\
&=&
\frac{g\phi_0^2 V_{BB-\kt} }{4\okt{}}\sum_{n=0}^\infty \frac{(-it)^n}{n!}\sum_{m=0}^{n-1}Q_1^{m}   (Q_1+\okt{})^{n-m-1} \vac_0\nonumber\\
&=&
\frac{g\phi_0^2 V_{BB-\kt} }{4\okt{}^2}e^{-iQ_1t}\left(e^{-i\okt{} t}-1
\right)\vac_0=\frac{\phi_0^2 \Delta_{\kt B} }{4\sQ}e^{-iQ_1t}\left(e^{-i\okt{} t}-1
\right)\vac_0.
\nonumber
\eea

Again, there is a corresponding contribution in which the free part of the kink Hamiltonian acts on the correction to the state
\beq
e^{-iH\p_2 t}|\kt\rangle_1^{\L 20}=
e^{-iQ_1 t}\frac{\phi^2_0}{4\sQ}\Delta_{-\kt,B}\vac_0.
\eeq
Adding these two contributions one finds the 0-meson, $\phi_0^2$ term in the evolved wave packet
\beq
e^{-iH\p t}|\kt\rangle \supset e^{-i(Q_1+\okt{}) t}\frac{\phi^2_0}{4\sQ}\Delta_{-\kt,B}\vac_0=e^{-i(Q_1+\okt{})t}|\kt\rangle^{\L 20}_1.
\eeq
In principle, when folding this into a wave packet, the same argument implies that this component moves via a rigid translation.  

However, the $\Delta_{-\kt,B}$ contains an $x$ integral which is supported near the kink and so this term vanishes at time $t=0$ when folded into the wave packet.  More generally it is
\bea
\pin{\kt} \alpha_\kt e^{-i(Q_1+\okt{})t}|\kt\rangle^{\L 20}_1&=&\pin{\kt}
2\sigma\sqrt{\pi}\mb_{\kt}e^{-\sigma^2\left(\kt-k_0\right)^2}e^{i(k_0-\kt)x_0}
e^{-i(Q_1+\okt{}) t}\frac{\phi^2_0}{4\sQ}\Delta_{-\kt,B}\vac_0\nonumber\\
&&\hspace{-3cm}=\frac{\sigma\sqrt{\pi}\phi^2_0}{2\sQ}e^{-i(Q_1+\ok{0})t}\int dx \g^\prime_B(x) \pin{\kt}\mb_\kt \g_{-\kt}(x) e^{-\sigma^2\left(\kt-k_0\right)^2}e^{i(k_0-\kt)x_t}\vac_0\nonumber\\
&&\hspace{-3cm}=\frac{\sigma\sqrt{\pi}\phi^2_0}{2\sQ}e^{-i(Q_1+\ok{0})t}\int dx \g^\prime_B(x) \pin{\kt}e^{i\kt x}e^{-\sigma^2\left(\kt-k_0\right)^2}e^{i(k_0-\kt)x_t}\vac_0\nonumber\\
&&\hspace{-3cm}=\frac{\phi^2_0}{4\sQ}e^{-i(Q_1+\ok{0})t}\int dx \g^\prime_B(x) e^{ik_0 x} e^{-\left(x_t-x\right)^2/(4\sigma^2)}\vac_0.
\eea
The Gaussian term $e^{-\left(x_t-x\right)^2/(4\sigma^2)}$ has support at $x\sim x_t$, with a width of $\sigma$.  On the other hand $\g_B\p(x)$ has support at $x=0$, with a width of $1/m$.  As a result, their product is exponentially suppressed unless $x_t$ is close to zero.  In other words, this term only turns on when the meson wave packet approaches the kink.  We conclude that the meson wave packet moves without deformation, apart from the usual spreading, until it comes within either its width $\sigma$ or the kink width $1/m$ of the kink, at which time corrections such as this one appear.

\subsubsection{$\phi_0$ Terms}

Let us next turn to the $(m,n)=(1,3)$ terms with one power of $\phi_0$ and three mesons.  Again there are two, the first of which arises from the subleading evolution operator acting on the leading meson state.  This uses the term
\beq
H\p_3\supset \frac{g\phi_0}{2} \ppink{2} V_{Bk_1k_2}\Bd{1}\Bd{2}\label{h31}
\eeq
leading to
\bea
e^{-iH\p t}\Big|_{O(g)}|\kt\rangle_0&\supset& \frac{g\phi_0}{2} \ppink{2} V_{Bk_1k_2}\sum_{n=0}^\infty \frac{(-it)^n}{n!}\\
&&\times \sum_{m=0}^{n-1}(Q_1+\okt{}+\ok{1}+\ok{2})^m  \Bd{1}\Bd{2} (Q_1+\omega_{\kt})^{n-m-1}|\kt\rangle_0\nonumber\\
&=& \frac{g\phi_0}{2} e^{-i(Q_1+\omega_{\kt})t} \ppink{2} \frac{V_{Bk_1k_2}}{\ok{1}+\ok{2}} \left(e^{-i(\ok{1}+\ok{2})t}- 1\right)|\kt;k_1k_2\rangle_0.\nonumber
\eea
Using the identity
\bea
V_{Bk_1k_2}&=&\int dx \V3 \g_B(x) \g_{k_1}(x)\g_{k_2}(x)\\
&=&\frac{1}{g\sQ}\int dx \V2 (\g_{k_1}(x)\g\p_{k_2}(x)+\g_{k_2}(x)\g\p_{k_1}(x))\nonumber\\
&=&\frac{1}{g\sQ}\int dx (\partial_x(\g\p_{k_1}(x)\g\p_{k_2}(x))+\ok{1}^2\g_{k_1}(x)\g\p_{k_2}(x)+\ok{2}^2\g_{k_2}(x)\g\p_{k_1}(x))\nonumber\\
&=&\frac{(\ok{1}^2-\ok{2}^2)\Delta_{k_1k_2}}{g\sQ}\nonumber
\eea
the contribution simplifies to
\beq
e^{-iH\p t}\Big|_{O(g)}|\kt\rangle_0
\supset
\frac{\phi_0}{2\sQ} e^{-i(Q_1+\omega_{\kt})t} \ppink{2} {{(\ok{1}-\ok{2})\Delta_{k_1k_2}}{}}{} \left(e^{-i(\ok{1}+\ok{2})t}- 1\right)|\kt;k_1k_2\rangle_0.
\eeq
Again, there is a second contribution in which the free evolution operator acts on the correction to the state
\bea
e^{-iH\p_2 t}|\kt\rangle^{\L 13}_1&=&e^{-iH\p_2 t}\left[\frac{\phi_0}{2\sQ}\ppink{2}\Delta_{k_1k_2}(\ok 2-\ok 1)|\kt;k_1k_2\rangle_0\right]\label{c13b}\\
&=&\frac{\phi_0}{2\sQ}e^{-i(Q_1+\okt{}) t}\ppink{2}\Delta_{k_1k_2}(\ok 2-\ok 1)e^{-i(\ok{1}+\ok{2}) t}|\kt;k_1k_2\rangle_0.\nonumber
\eea

Combining these contributions, one arrives at the total 3-meson, one power of $\phi_0$ piece of the state
\beq
e^{-iH\p t}|\kt\rangle \supset \frac{\phi_0}{2\sQ}e^{-i(Q_1+\okt{}) t}\ppink{2}\Delta_{k_1k_2}(\ok 2-\ok 1)|\kt;k_1k_2\rangle_0 =e^{-i(Q_1+\okt{})t}|\kt\rangle^{\L 13}_1.
\eeq
Again, we see that this component of the state evolves via a phase rotation $e^{-i(Q_1+\okt{}) t}$.  The argument used above at leading order again implies that, when folded into the wave packet, this contribution evolves via rigid translation, as desired.  

Physically this term is just an undressed meson wave packet in the presence of the leading quantum correction to the kink.  As the kink is far from the meson, it was to be expected that its quantum correction does not affect the propagation of the meson.

The last descendant term that may arise consists of one power of the zero mode $\phi_0$ and one meson.   Recall that we found such a term in the initial condition which fixes the distance between the kink and the meson wave packet.  Again, the first contribution arises from the quantum correction to the evolution operator, now using
\beq
H\p_3\supset \frac{g\phi_0}{2} \ppink{2} \frac{V_{Bk_1k_2}}{\ok{2}}\Bd{1}B_{-k_2}
\eeq
one finds the contribution
\bea
e^{-iH\p t}\Big|_{O(g)}|\kt\rangle_0&\supset& \frac{g\phi_0}{2} \ppink{2} \frac{V_{Bk_1k_2}}{\ok{2}}\sum_{n=0}^\infty \frac{(-it)^n}{n!}\\
&&\times \sum_{m=0}^{n-1}(Q_1+\ok{1})^m  \Bd{1}B_{-k_2} (Q_1+\omega_{\kt})^{n-m-1}|\kt\rangle_0\nonumber\\
&=& \frac{g\phi_0}{2\okt{}} e^{-i(Q_1+\omega_{\kt})t} \ppin{k} \frac{V_{B,k,-\kt}}{\ok{}-\okt{}} \left(e^{-i(\ok{}-\okt{})t}- 1\right)|k\rangle_0\nonumber
\\
&=& \frac{\phi_0}{2\sQ} e^{-i(Q_1+\omega_{\kt})t} \ppin{k} \left(1+\frac{\ok{}}{\okt{}}\right)\Delta_{k,-\kt} \left(e^{-i(\ok{}-\okt{})t}- 1\right)|k\rangle_0.\nonumber
\eea
Adding this to
\bea
e^{-iH\p_2 t}|\kt\rangle^{\L 11}_1&=&e^{-iH\p_2 t}\left[\frac{\phi_0}{2\sQ}\ppin{k}\Delta_{-\kt,k}\left(1+\frac{\ok{}}{\okt{}}\right)|k\rangle_0\right]\label{c13b}\\
&=&\frac{\phi_0}{2\sQ}e^{-i(Q_1+\okt{}) t}\ppin{k}\Delta_{-\kt,k}\left(1+\frac{\ok{}}{\okt{}}\right)e^{-i(\ok{}-\okt{}) t}|k\rangle_0\nonumber
\eea
one finds the total one-meson contribution
\beq
e^{-iH\p t}|\kt\rangle \supset \frac{\phi_0}{2\sQ}e^{-i(Q_1+\okt{}) t}\ppin{k}\Delta_{-\kt,k}\left(1+\frac{\ok{}}{\okt{}}\right)|k\rangle_0
 =e^{-i(Q_1+\okt{})t}|\kt\rangle^{\L 11}_1.
\eeq
Again the fact that the full evolution yields a factor of $e^{-i(Q_1+\okt{})t}$ implies that even this term, although it is perhaps the most unexpected in our wave packet as it does not arise from the action of the dressed creation operator on the kink ground state, nonetheless evolves via a rigid translation of the meson wave packet towards the kink.  

Recall that this term enforces that, in each kink-position eigenstate component of the momentum eigenstate, the meson wave packet is at the same distance from the kink.  One thus learns that this same distance evolves via rigid motion in each of these kink-position eigenstates.  In other words, wherever the kink may lie, the meson wave packet moves towards it rigidly and with the same speed.

In principle there are other contributions with one meson.  One potential source of these is the interaction
\beq
H\p_3\supset \frac{g\phi_0}{2}  {V_{\I  B}}.
\eeq
Via an argument similar to those above, it was shown in Appendix A of Ref.~\cite{me2loop} that $V_{\I  B}=0$, and so there is no such contribution.  Similarly a contribution could arise from the $V_{BBB}$ term in $H\p_3$ combined with the $\pi_0^2/2$ in $H\p_2$, however we have shown above that $V_{BBB}=0$ and so there is also no such contribution.

\subsection{Subleading Order: Primaries}

\subsubsection{The Four-Meson Sector}

The four-meson sector describes the traveling meson together with three-meson virtual excitations around the kink.  As the virtual excitation cloud is localized around the kink, one expects it not to interact with the virtual meson.  To check that this intuition is correct, we consider the two usual contributions.  The first consists of the leading quantum correction to the evolution operator corresponding to 
\beq
H\p_3\supset \frac{g}{6} \ppink{3} V_{k_1k_2k_3}\Bd{1}\Bd{2}\Bd{3}.
\eeq
This leads to the correction
\bea
e^{-iH\p t}\Big|_{O(g)}|\kt\rangle_0&\supset& \frac{g}{6} \ppink{3} {V_{k_1k_2k_3}}{}\sum_{n=0}^\infty \frac{(-it)^n}{n!}\\
&&\times \sum_{m=0}^{n-1}(Q_1+\okt{}+\ok{1}+\ok{2}+\ok{3})^m  \Bd{1}\Bd{2}\Bd{3} (Q_1+\okt{})^{n-m-1}|\kt\rangle_0\nonumber\\
&=& \frac{g}{6} e^{-i(Q_1+\omega_{\kt})t} \ppin{k} \frac{V_{k_1k_2k_3}}{\ok{1}+\ok{2}+\ok{3}} \left(e^{-i(\ok{1}+\ok{2}+\ok{3})t}- 1\right)|\kt;k_1k_2k_3\rangle_0.\nonumber
\eea

On the other hand, the contribution from the free evolution operator acting on the corrected state is
\bea
e^{-iH\p_2 t}|\kt\rangle^{\L 04}_1&=&e^{-iH\p_2 t}\left[-\frac{g}{6}\ppink{3} \frac{V_{k_1k_2k_3}}{\ok1+\ok2+\ok3}|\kt;k_1k_2k_3\rangle_0\right]\\
&=&-\frac{g}{6} e^{-i(Q_1+\omega_{\kt})t} \ppink{3} \frac{V_{k_1k_2k_3}}{\ok{1}+\ok{2}+\ok{3}}e^{-i(\ok{1}+\ok{2}+\ok{3})t}|\kt;k_1k_2k_3\rangle_0.\nonumber
\eea
Adding these two together, as expected one finds
\beq
e^{-iH\p t}|\kt\rangle \supset -\frac{g}{6} e^{-i(Q_1+\omega_{\kt})t} \ppink{3} \frac{V_{k_1k_2k_3}}{\ok{1}+\ok{2}+\ok{3}}|\kt;k_1k_2k_3\rangle_0
 =e^{-i(Q_1+\okt{})t}|\kt\rangle^{\L 04}_1.
\eeq
Thus we have confirmed that even in this sector, in which the meson wave packet arrives at the corrected kink from far away, the meson wave packet moves rigidly.

\subsubsection{The Two-Meson Sector}

This sector is a bit more complicated then the others, as the free kinetic term $\pi_0^2/2$ in $H\p_2$ also contributes and also there are three terms in $|\kt\rangle^{\L 02}_1$
\bea
|\kt\rangle^{\L02}_1&=&|\kt\rangle^{\L02A}_1+|\kt\rangle^{\L02B}_1+|\kt\rangle^{\L02C}_1\\
|\kt\rangle^{\L02A}_1&=&- \frac{g}{2}\ppin{k} \frac{V_{\I k}}{\ok{}}|\kt;k\rangle_0
\nonumber\\
|\kt\rangle^{\L02B}_1&=& -\frac{1}{2\sQ}\ppin{k} {{\Delta_{kB} }{}}{}|\kt;k\rangle_0
\nonumber\\
|\kt\rangle^{\L02C}_1&=& \frac{g V^{(3)}_-}{4\okt{}}\int dx 
\ppink{2} \frac{\g_{k_1}(x) \g_{k_2}(x) \g_{-\kt}(x)}{\okt{}-\ok1-\ok2} |k_1k_2\rangle_0 .
\nonumber
\eea

Let us start with the interaction term
\beq
H\p_3\supset \frac{g}{4} \ppink{3} \frac{V_{k_1k_2k_3}}{\ok{3}}\Bd{1}\Bd{2}B_{-k_3}.
\eeq
This is annihilated by $\pi_0$ and so the usual formulas can be applied
\bea
e^{-iH\p t}\Big|_{O(g)}|\kt\rangle_0&\supset& \frac{g}{4} \ppink{3} \frac{V_{k_1k_2k_3}}{\ok{3}}\sum_{n=0}^\infty \frac{(-it)^n}{n!}\\
&&\times \sum_{m=0}^{n-1}(Q_1+\ok{1}+\ok{2})^m  \Bd{1}\Bd{2}B_{-k_3}(Q_1+\okt{})^{n-m-1}|\kt\rangle_0\nonumber\\
&=& \frac{g}{4} e^{-i(Q_1+\omega_{\kt})t} \ppink{2} \frac{V_{k_1k_2-\kt}}{\ok{1}+\ok{2}-\okt{}} \left(e^{-i(\ok{1}+\ok{2}-\okt{})t}- 1\right)|;k_1k_2\rangle_0.\nonumber
\eea
The corresponding term arising from the leading evolution operator is
\bea
e^{-iH\p_2 t}|\kt\rangle^{\L 02C}_1&=&e^{-iH\p_2 t}\left[\frac{g V^{(3)}_-}{4\okt{}}\int dx 
\ppink{2} \frac{\g_{k_1}(x) \g_{k_2}(x) \g_{-\kt}(x)}{\okt{}-\ok1-\ok2} |k_1k_2\rangle_0 \right]\\
&&\hspace{-2cm}=\frac{g V^{(3)}_-}{4\okt{}} e^{-i(Q_1+\omega_{\kt})t}\int dx 
\ppink{2} \frac{\g_{k_1}(x) \g_{k_2}(x) \g_{-\kt}(x)}{\okt{}-\ok1-\ok2}e^{-i(\ok{1}+\ok{2}-\okt{})t} |k_1k_2\rangle_0 .\nonumber
\eea
Adding these two terms we arrive at
\bea
e^{-iH\p t}|\kt\rangle
 &\supset&e^{-i(Q_1+\okt{})t}|\kt\rangle^{\L 02C}_1+C_\kt\label{ck}\\
C_\kt&=&
\frac{g }{4\okt{}} e^{-iQ_1t}\int dx \left(V^{(3)}_--\V3
\right)\nonumber\\
&&\times
\ppink{2} \frac{\g_{k_1}(x) \g_{k_2}(x) \g_{-\kt}(x)}{\okt{}-\ok1-\ok2}\left(e^{-i(\ok{1}+\ok{2})t}-e^{-i\okt{}t}\right) |k_1k_2\rangle_0 .
\nonumber
\eea

The cancellation that we have always seen between these two terms, annihilating the term that has the on-shell dispersion relation for the virtual particles, does not quite work here.  Instead a remainder $C_\kt$ remains.  The problem is that one expression has the full $V_{k_1k_2-\kt}$ arising from the kink Hamiltonian, while the other has only $V^{(3)}_-$ arising from the left vacuum Hamiltonian.

Let us fold $C_\kt$ into the wave packet, to see if this term is present in the evolved state.  This yields
\bea
\pin{\kt} \alpha_{\kt}C_{\kt}&=&
\frac{g\sigma\sqrt{\pi}e^{-iQ_1t}}{2}
\int dx \left(V^{(3)}_--\V3
\right)\ppink{2} \g_{k_1}(x) \g_{k_2}(x)|k_1k_2\rangle_0
 \nonumber\\
&&\hspace{-2cm}\times
 \pin{\kt}
\frac{\mb_{\kt}}{\okt{}}e^{-\sigma^2\left(\kt-k_0\right)^2}e^{i(k_0-\kt)x_0}\frac{ \g_{-\kt}(x)}{\okt{}-\ok1-\ok2}\left(e^{-i(\ok{1}+\ok{2})t}-e^{-i\okt{}t}\right).
\eea
Let us look more closely at the second line.  It depends on $k_1$, $k_2$ and $x$.  The integrand does not have a pole when the on-shell condition $\ok{1}+\ok{2}=\okt{}$ is satisfied because at that point the term in parentheses also vanishes linearly, leaving a finite on-shell limit for the integrand.  

Again expanding the dispersion relation for $\okt{}$, we may rewrite this second line as
\beq
\pin{\kt}
\frac{\mb_{\kt}}{\okt{}}e^{-\sigma^2\left(\kt-k_0\right)^2}\frac{ \g_{-\kt}(x)}{\okt{}-\ok1-\ok2}\left(e^{-i(\ok{1}+\ok{2})t}e^{i(k_0-\kt)x_0}-e^{-i\ok{0}t}e^{i(k_0-\kt)x_t}\right).
\eeq
Thus we see that the phase of $\g_{-\kt}(x)$ must vary, with respect to $\kt$, with a derivative of about $x_0$ if the first term is to contribute or $x_t$ if the second is to contribute.  This means that this term is exponentially suppressed unless $x\sim x_0$ or $x\sim x_t$.  On the other hand, the $V^{(3)}_--\V3$ factor is exponentially suppressed if $x\ll 0$.  Thus both terms can be nonvanishing only if $x_0\gtrsim 0$, which it is not, or if $x_t\gtrsim 0$, which occurs once the meson wave packet approaches the kink. 

In conclusion, the correction term $C_\kt$ vanishes, when folded into the meson wave packet, until the meson is within a distance of order $1/m$ or within a distance of order $\sigma$ of the kink.  Before this time, $|\kt\rangle^{\L 02C}_1$ evolves via a multiplication by the phase $e^{-i(Q_1+\okt{})t}$ and so, when folded into the wave packet, evolves via a rigid translation.  This resolves the puzzle of how a wave packet that was constructed using eigenstates of the left vacuum Hamiltonian $H^\L$ may transform rigidly under $H\p$ evolution, the difference between the two eigenstates is in the kernel of the integral weighted by $\alpha_\kt$.

Let us now turn to the next interaction term
\beq
H\p_3\supset \frac{g}{2} \ppin{k} V_{\I  k}\Bd{}
\eeq
which leads to the evolution
\bea
e^{-iH\p t}\Big|_{O(g)}|\kt\rangle_0&\supset& \frac{g}{2} \ppin{k} {V_{\I  k}}{}\sum_{n=0}^\infty \frac{(-it)^n}{n!}\\
&&\times \sum_{m=0}^{n-1}(Q_1+\okt{}+\ok{})^m  \Bd{}(Q_1+\okt{})^{n-m-1}|\kt\rangle_0\nonumber\\
&=& \frac{g}{2} e^{-i(Q_1+\omega_{\kt})t} \ppin{k} \frac{V_{\I-\kt}}{\ok{}} \left(e^{-i\ok{}t}- 1\right)|\kt;k\rangle_0.\nonumber
\eea
Adding
\bea
e^{-iH\p_2 t}|\kt\rangle^{\L 02C}_1&=&e^{-iH\p_2 t}\left[- \frac{g}{2}\ppin{k} \frac{V_{\I k}}{\ok{}}|\kt;k\rangle_0 \right]\\
&&\hspace{-2cm}=-\frac{g}{2} e^{-i(Q_1+\omega_{\kt})t}\ppin{k} \frac{V_{\I k}}{\ok{}}e^{-i\ok{} t}|\kt;k\rangle_0 \nonumber
\eea
as usual leads to
\beq
e^{-iH\p t}|\kt\rangle \supset e^{-i(Q_1+\okt{})t}|\kt\rangle^{\L 02C}_1.
\eeq

The final contribution is a bit different.   Consider the interaction
\beq
H\p_3\supset \frac{g\phi_0^2}{2} \ppin{k} V_{BBk}\Bd{}.
\eeq
This interaction can, despite the $\phi_0^2$ factor, lead to a primary in the evolution operator because $H\p_2$ contains a $\pi_0^2/2$ term.  The corresponding contribution is
\bea
e^{-iH\p t}\Big|_{O(g)}|\kt\rangle_0&\supset& \frac{g}{2} \ppin{k} {V_{BB k}}{}\sum_{n=0}^\infty \frac{(-it)^n}{n!}\label{primev}\\
&&\hspace{-3cm}\times \sum_{m=0}^{n-1}\left(Q_1+\okt{}+\ok{}+\frac{\pi_0^2}{2}\right)^m  \phi_0^2\Bd{}(Q_1+\okt{})^{n-m-1}|\kt\rangle_0\nonumber\\
&&\hspace{-3cm}=-\frac{g}{2} \ppin{k} {V_{BB k}}{}\sum_{n=0}^\infty \frac{(-it)^n}{n!}\sum_{m=0}^{n-1}m\left(Q_1+\okt{}+\ok{}\right)^{m-1}(Q_1+\okt{})^{n-m-1}|\kt;k\rangle_0\nonumber\\
&&\hspace{-3cm}=-\frac{g}{2} \ppin{k} {V_{BB k}}{}\frac{\partial}{\partial\ok{}}\left[\sum_{n=0}^\infty \frac{(-it)^n}{n!}\sum_{m=0}^{n-1}\left(Q_1+\okt{}+\ok{}\right)^{m}(Q_1+\okt{})^{n-m-1}\right]|\kt;k\rangle_0\nonumber\\
&&\hspace{-3cm}= \frac{1}{2\sQ} e^{-i(Q_1+\omega_{\kt})t} \ppin{k} {\Delta_{k B}}\left({i\ok{}t e^{-i\ok{}t}+ e^{-i\ok{}t}-1}{}\right)|\kt;k\rangle_0.\nonumber
\eea
The linear growth in $t$ may look worrying.

The other contributions arise from the free evolution operator acting on the excited state.  There are two such contributions.  One is simply
\bea
e^{-iH\p_2 t}|\kt\rangle^{\L 02B}_1&=&e^{-iH\p_2 t}\left[ -\frac{1}{2\sQ}\ppin{k} {{\Delta_{kB} }{}}{}|\kt;k\rangle_0 \right]\\
&&\hspace{-2cm}=-\frac{1}{2\sQ} e^{-i(Q_1+\omega_{\kt})t}\ppin{k} {\Delta_{kB}}{}e^{-i\ok{} t}|\kt;k\rangle_0 \nonumber
\eea
and it cancels the second term in the parenthesis in Eq.~(\ref{primev}).  The other uses the kink kinetic term $\pi_0^2/2$
\bea
e^{-iH\p_2 t}|\kt\rangle^{\L 22}_1&=&\sum_{n=0}^\infty \frac{(-it)^n}{n!}\left[
-\frac{1}{2\sQ}\ppin{k}
\left(
Q_1+\ok{}+\okt{}+\frac{\pi_0^2}{2}
\right)^n
\Delta_{kB}\ok{} \phi^2_0|\kt;k\rangle_0
\right]\nonumber\\
&=&-\frac{it}{2\sQ}\sum_{n=1}^\infty \frac{(-it)^{n-1}}{(n-1)!}\left[
\ppin{k}
\left(
Q_1+\ok{}+\okt{}
\right)^{n-1}
\Delta_{kB}\ok{} |\kt;k\rangle_0
\right]\nonumber\\
&=&-\frac{1}{2\sQ}
\ppin{k}
\Delta_{kB}i\ok{}t e^{-i(Q_1+\ok{}+\okt{})t} |\kt;k\rangle_0
\eea
and it cancels the first term in the parenthesis in Eq.~(\ref{primev}).

Adding all three contributions together, one arrives at
\bea
e^{-iH\p t}|\kt\rangle
 \supset -\frac{1}{2\sQ} e^{-i(Q_1+\omega_{\kt})t} \ppin{k} {\Delta_{k B}}|\kt;k\rangle_0
 &=&e^{-i(Q_1+\okt{})t}|\kt\rangle^{\L 02B}_1.
\eea
Again, when folded into the wave packet this ensures that even the two-meson states move along with the wave packet.  This is nontrivial of course, since if the two mesons had been on-shell they would move more slowly than the single meson.

\subsubsection{The No-Meson Sector}

Finally we turn our attention to the no-meson sector of the primary coefficients.  There are two such terms in the initial condition
\bea
|\kt\rangle^{\L00}_1&=&|\kt\rangle^{\L00A}_1+|\kt\rangle^{\L00B}_1\\
|\kt\rangle^{\L00A}_1&=&-\frac{\Delta_{-\kt,B}}{4\sQ\okt{}}
\vac_0\hsp
|\kt\rangle^{\L00B}_1=\frac{g V^{(3)}_-}{4\okt{}^2}\int dx\I(x)\g_{-\kt}(x)\vac_0.\nonumber
\eea

The only interaction which contributes to the second term is
\beq
H\p_3\supset \frac{g}{4} \ppin{k} \frac{V_{\I  k}}{\ok{}}B_{-k}
\eeq
leading to
\bea
e^{-iH\p t}\Big|_{O(g)}|\kt\rangle_0&\supset& \frac{g}{4} \ppin{k} \frac{V_{\I k}}{\ok{3}}\sum_{n=0}^\infty \frac{(-it)^n}{n!}\sum_{m=0}^{n-1}Q_1^m  B_{-k}(Q_1+\okt{})^{n-m-1}|\kt\rangle_0\nonumber\\
&=&\frac{gV_{\I -\kt}}{4\okt{}^2} e^{-iQ_1t}  \left(e^{-i\okt{}t}-1\right)\vac_0.
\eea
Adding this to
\bea
e^{-iH\p_2 t}|\kt\rangle^{\L 00B}_1&=&e^{-iH\p_2 t}\left[\frac{g V^{(3)}_-}{4\okt{}^2}\int dx\I(x)\g_{-\kt}(x)\vac_0 \right]\\
&=&\frac{g V^{(3)}_-}{4\okt{}^2} e^{-iQ_1 t} \int dx\I(x)\g_{-\kt}(x)\vac_0 \nonumber
\eea
we find
\bea
e^{-iH\p t}|\kt\rangle
 &\supset&e^{-i(Q_1+\okt{})t}|\kt\rangle^{\L 00B}_1+D_\kt \label{zev}\\
D_\kt&=&
\frac{g }{4\okt{}^2} e^{-iQ_1t}\int dx \left(V^{(3)}_--\V3
\right)\I(x)\g_{-\kt}(x)\left(1-e^{-i\okt{}t}\right) |k_1k_2\rangle_0 .
\nonumber
\eea

The remainder $D_\kt$ is very similar to the $C_\kt$ in Eq.~(\ref{ck}).  As a result, the same argument used above to show that $C_\kt$ vanishes when folded into the wave packet also applies to $D_\kt$.  This an important consistency check.  Recall that the zero-meson sector even in the initial condition was annihilated by the folding into the wave packet, and so there was no zero-meson piece in the initial state.  Now in (\ref{zev}) we found that the evolution consists of two terms.  The first, due to the factor of $e^{-i(Q_1+\okt{})t}$, implies that the zero-meson piece of the wave packet is translated rigidly.  However, when folded into the wave packet this piece is zero, and so there is nothing to translate rigidly.  The $D_\kt$ piece on the other hand is, as argued above in the case of $C_\kt$, generated only when the wave packet arrives within a distance of $\sigma$ or $1/m$ of the kink, and so it results from the kink-meson scattering.  Thus, we find that the meson wave packet indeed is rigidly translated before reaching the kink, despite the fact that the initial condition was defined using the vacuum Hamiltonian, which is a truncation of the true dynamics, while the evolution is performed using the kink Hamiltonian.


Next let us turn to the first term $|\kt\rangle^{\L 00A}$.     The first contribution arises from the interaction
\beq
H\p_3\supset \frac{g\phi_0^2}{4} \ppin{k} \frac{V_{BBk}}{\ok{}}B_{-k}
\eeq
which leads to
\bea
e^{-iH\p t}\Big|_{O(g)}|\kt\rangle_0&\supset& \frac{g}{4}e^{-iQ_1t} \ppin{k} \frac{V_{BB k}}{\ok{}}\sum_{n=0}^\infty \frac{(-it)^n}{n!} \sum_{m=0}^{n-1}\left(\frac{\pi_0^2}{2}\right)^m  \phi_0^2 B_{-k}\okt{}^{n-m-1}|\kt\rangle_0\nonumber\\
&=&-\frac{gV_{BB -\kt}}{4\okt{}} e^{-iQ_1t}{}{}\sum_{n=2}^\infty \frac{(-it)^n}{n!}\okt{}^{n-2}\vac_0\nonumber\\
&=&-\frac{\Delta_{-\kt B}}{4\sQ\okt{}} e^{-iQ_1t}{}\left(e^{-i\okt{}t}-1+it\okt{}
\right)
\vac_0.\label{nomes}
\eea
Other contributions arise from the free evolution operator acting on the excited state.  The first is
\bea
e^{-iH\p_2 t}|\kt\rangle^{\L 00A}_1&=&e^{-iH\p_2 t}\left[-\frac{\Delta_{-\kt,B}}{4\sQ\okt{}}
\vac_0 \right]
=-\frac{\Delta_{-\kt,B}}{4\sQ\okt{}}e^{-iQ_1 t}\vac_0
\eea
which cancels the second term in parenthesis in Eq.~(\ref{nomes}).  The other uses the kink kinetic term $\pi_0^2/2$
\bea
e^{-iH\p_2 t}|\kt\rangle^{\L 20}_1&=&
\sum_{n=0}^\infty \frac{(-it)^n}{n!}\left(\frac{\pi_0^2}{2}\right)^n
\frac{\phi^2_0}{4\sQ}\Delta_{-\kt,B}\vac_0=it\frac{\Delta_{-\kt,B}}{4\sQ}e^{-iQ_1 t}\vac_0
\eea
and cancels the last term in Eq.~(\ref{nomes}).  In the end, only the first term in (\ref{nomes}) remains
\bea
e^{-iH\p t}|\kt\rangle
 \supset -\frac{\Delta_{-\kt B}}{4\sQ\okt{}} e^{-i(Q_1+\okt{})t}{}
\vac_0
 &=&e^{-i(Q_1+\okt{})t}|\kt\rangle^{\L 00A}_1.
\eea
For the last time, a change in the phase shift, folded into the wave packet, yields a rigid translation.  We thus have completed our demonstration that our prescription for the leading correction to the wave packet for a meson incident on a kink evolves, under the full kink Hamiltonian, by a rigid translation with no deformations, even to the quantum corrections, before the meson wave packet physically overlaps with the kink.

\section{Subleading Corrections} \label{ggsez}

The main motivation for the present work is to prepare for a treatment of elastic kink-meson scattering, whose amplitude is expected to be of order $O(g^2)$.  One potential contribution to this will be an $O(g^2)$ correction to the asymptotic state proportional to $|-\kt\rangle_0$.  To see if such a contribution is present, we need to calculate the $O(g^2)$ correction $|\kt\rangle^{\L}_2$ to the one-meson part of $|\kt\rangle^\L$.  We will evaluate the amplitude using the reduced norm of Ref.~\cite{menorm}, which only requires the primary part $|\kt\rangle^{\L 01}_2$.  We will find that there is no $|-\kt\rangle_0$ term in $|\kt\rangle^\L_2$, or more precisely that this term can and should be set to zero, and so there will be no corresponding correction to elastic kink-meson scattering.  This question was the main motivation for the present work.

\subsection{One Meson and Two Zero Modes}

To find $|\kt\rangle^{\L 01}_2$, first we need to find the term $|\kt\rangle^{\L 21}_2$ with two zero modes.  As this is a descendant, it is determined entirely by translation invariance $P\p|\kt\rangle^\L=0$.  This condition constrains $\pi_0|\kt\rangle^{\L 21}_2$ 
\bea
\pi_0|\kt\rangle^{\L 21}_2
&=&-\frac{1}{\sQ}P|\kt\rangle^\L_1\Large|_{m=n=1}.
\eea
Which terms in $|\kt\rangle^\L_1$ may contribute?  We recall that all of the zero-meson terms are annihilated when folded into the wave packet as they contain $\Delta_{\kt B}$, $V_{\I\kt}$ or its vacuum Hamiltonian analogue which are themselves annihilated.  This means that their contributions to $|\kt\rangle^\L_2$ will also be annihilated when folded into the wave packet,  and so we will simply drop them.

This leaves four contributions.  The $m=n=1$ terms of each are denoted with the $\supset$ symbol
\bea
P|\kt\rangle^{\L11}_1
&\supset&
-\frac{i}{2}\ppink{2}\Delta_{k_1k_2}\left(1+\frac{\omega_{k_1}}{\omega_{k_2}}\right)B^\ddag_{k_1}B_{-k_2}
\left[
\frac{\phi_0}{2\sQ}\ppin{k}\Delta_{-\kt,k}\left(1+\frac{\ok{}}{\okt{}}\right)|k\rangle_0
\right]
\nonumber\\
&=&-\frac{i\phi_0}{4\okt{}\sQ}\ppin{k}\left[\ppin{k\p}\frac{\Delta_{-\kt,-k\p}\Delta_{kk\p}}{\okp{}}\left(\okt{}+{\okp{}}{}\right)\left(\ok{}+{\okp{}}{}\right)\right]|k\rangle_0
\eea
\bea
P|\kt\rangle^{\L22}_1
&\supset&
\ppin{k}\Delta_{kB}\pi_0\frac{B_{-k}}{2\omega_k}
\left[
-\frac{\phi^2_0}{2\sQ}\ppin{k\p}\Delta_{k\p  B}\okp{} |\kt;k\p\rangle_0
\right]
\nonumber\\
&=&\frac{i\phi_0}{2\sQ}\ppin{k}\left[
\frac{\ok{}}{\okt{}}\Delta_{-\kt B}\Delta_{kB}|;k\rangle_0
+
|\Delta_{kB}|^2|\kt\rangle_0
\right]
\eea
\bea
P|\kt\rangle^{\L02}_1
&\supset&
\ppin{k}\Delta_{kB}i\phi_0\frac{B_{-k}}{2}
\left[
\frac{g V^{(3)}_-}{4\okt{}}\int dx 
\ppink{2} \frac{\g_{k_1}(x) \g_{k_2}(x) \g_{-\kt}(x)}{\okt{}-\ok1-\ok2} |k_1k_2\rangle_0\right.\nonumber\\
&&\left.
-\frac{g}{2}\ppin{k\p} {\left[\frac{V_{\I k\p}}{\okp{}}+\frac{\Delta_{k\p B}}{g\sQ}
\right]}{}|\kt;k\p\rangle_0
\right]
\nonumber\\
&=&
\frac{ig\phi_0 V^{(3)}_-}{4\okt{}}\ppin{k}\left[\ppin{k\p}
\Delta_{-k\p B}\int dx \frac{\g_{k}(x)\g_{k\p}(x)\g_{-\kt}(x)}{\okt{}-\ok{}-\okp{}}
\right]|;k\rangle_0
\nonumber\\
&&-\frac{ig\phi_0}{4}
\ppin{k} {\left[\frac{V_{\I k}}{\ok{}}+\frac{\Delta_{k B}}{g\sQ}
\right]}{}\left(\Delta_{-k B}|\kt\rangle_0+\Delta_{-\kt B}|;k\rangle_0
\right)
\eea
\bea
P|\kt\rangle^{\L13}_1
&\supset&
i\ppink{2}\Delta_{k_1k_2}\frac{B_{-k_1}B_{-k_2}}{4\omega_{k_2}}
\left[
\frac{\phi_0}{2\sQ}\ppinkp{2}\Delta_{k\p_1k\p_2}(\okp 2-\okp 1)|\kt;k\p_1k\p_2\rangle_0
\right]
\nonumber\\
&=&
-\frac{i\phi_0}{8\sQ}\left[\ppink{2}|\Delta_{k_1k_2}|^2 \frac{(\ok1-\ok2)^2}{\ok1\ok2}
\right]|\kt\rangle_0\nonumber\\
&&+\frac{i\phi_0}{4\okt{}\sQ}\ppin{k}\left[\ppin{k\p}\frac{\Delta_{-\kt ,-k\p}\Delta_{k\p k}}{\okp{}}(\okt{}-\okp{})
(\ok{}-\okp{})\right]|k\rangle_0.
\eea
Some of these terms will also vanish when folded into the wave packet.  For example, the $|;k\rangle_0$ term in $P|\kt\rangle^{\L22}_1$ is proportional to $\Delta_{-\kt B}$, which we have seen vanishes when folded as the $x$ integral is supported far from the origin where $\g_B(x)$ vanishes.  

What about $P|\kt\rangle^{\L02}_1$?  The argument above applies to the $\Delta_{-\kt B}$ term here as well, so that term will not contribute.  Let us next consider the term that is trilinear in $\g$, which is also of the form $|;k\rangle_0$.  When folded into the wave packet, the $\kt$ integral vanishes unless $x\sim x_0$.  Now, consider the $k\p$ integral.  This consists of an $\okp{}$ in the denominator which varies slowly far from the pole, and also $\g_{k\p}(x)$ in the numerator and a $\g_{-k\p}(y)$ in the $\Delta_{-k\p B}$.  What about the pole?   Now recall that in Eq.~(\ref{vh}), the vacuum Hamiltonian should use the vacuum field $\phi^\L$ and so the three normal mode factors in the residue are in fact plane waves, as can be seen in (\ref{init}).  The $x$ integration then implies that $k_1+k_2=\kt$ and so the pole is avoided.  This is only the second time that the vacuum field has been relevant in this note.
Therefore, the $\kt{}$-dependence in the denominator has little effect on the $k\p$ integral, whose integrand changes phase very rapidly with respect to $k\p$ as a result of the $e^{-ik\p x}$ in $\g_k\p(x)$.  This must be compensated by a phase in $\g_{-k\p}(y)$.  This requires $y\sim x$, which we recall is very large.  As a result the $\g^\prime_B(y)$ in $\Delta_{-k\p B}$ vanishes.  We conclude that this term will vanish when folded into the wave packet, and so we do not consider it further.  Only the $|\kt\rangle_0$ piece of the $P|\kt\rangle^{\L02}_1$ term may contribute.

In fact, we may also simplify the first and fourth terms when folded into wave packets.  Recall from Eq.~(\ref{i1}) that this folding allows us to replace $\Delta_{-\kt,k}$ with $-i\kt 2\pi\delta(\kt-k)$.  As a result of the $\delta$, the round parenthesis in the last line of $P|\kt\rangle^{\L13}_1$ vanish, and so this line does not contribute leaving
\beq
P|\kt\rangle^{\L13}_1\supset -\frac{i\phi_0}{8\sQ}\left[\ppink{2}|\Delta_{k_1k_2}|^2 \frac{(\ok1-\ok2)^2}{\ok1\ok2}
\right]|\kt\rangle_0 .
\eeq
On the other hand, it allows the integrals in $P|\kt\rangle^{\L11}_1$ to be performed, yielding
\bea
P|\kt\rangle^{\L11}_1
&\supset&-\frac{i\kt^2\phi_0}{\sQ}|\kt\rangle_0.\nonumber
\eea
Adding these to the $|\kt\rangle_0$ terms in $P|\kt\rangle^{\L22}_1$ and $P|\kt\rangle^{\L02}_1$ one arrives at
\bea
P|\kt\rangle^{\L}_1
&\supset&\frac{i\phi_0}{4\sQ}\left[-4\kt^2+
\ppin{k}
\left(|\Delta_{kB}|^2-g\sQ \frac{\Delta_{-kB} V_{\I k}}{\ok{}}
\right)\right.\\
&&\left.
-\frac{1}{2}\ppink{2}|\Delta_{k_1k_2}|^2 \frac{(\ok1-\ok2)^2}{\ok1\ok2}
\right]
|\kt\rangle_0\nonumber\\
&=&-\sqrt{Q_0}\pi_0|\kt\rangle^{\L 21}_2
.\nonumber
\eea
Thus we have found
\bea
|\kt\rangle^{\L 21}_2&=&\frac{\phi^2_0}{8Q_0}\left[-4\kt^2+
\ppin{k}
\left(|\Delta_{kB}|^2-g\sQ \frac{\Delta_{-kB} V_{\I k}}{\ok{}}
\right)\right.\\
&&\left.
-\frac{1}{2}\ppink{2}|\Delta_{k_1k_2}|^2 \frac{(\ok1-\ok2)^2}{\ok1\ok2}
\right]
|\kt\rangle_0.\nonumber
\eea
Note that the only momentum which appears is $k=\kt$.  This was also the case  at $O(g)$ and is to be expected in general.  It reflects the fact that the meson momentum itself is conserved far from the kink, as momentum cannot be exchanged with the kink from far away.    This fact distinguishes the asymptotic states defined in this paper from true eigenstates of the kink Hamiltonian, calculated in Ref.~\cite{menorm}, which also contain $|-\kt\rangle_0$, even after being folded into a distant wave packet, reflecting the fact that the time-independent states contain a component which is reflected from the kink.  Such elastic meson-kink scattering will be considered in the near future.

\subsection{One Meson and No Zero Modes}

Finally we are ready to evaluate $|\kt\rangle^{\L 01}_2$.  Recall (\ref{init}) that the state $|\kt\rangle^\L_0+|\kt\rangle^\L_1$, when folded into the wave packet, is equal to 
\bea
\left(B^\ddag_{\kt}-
\frac{g V^{(3)}_-}{4\okt{}}
\pin{k} \frac{\mb_{k} \mb_{\kt-k} \mb_{-\kt}}{\okt{}-\ok{}-\omega_{\kt-k}} B^\ddag_kB^\ddag_{\kt-k}\right)\vac
-\frac{i\kt \phi_0}{\sQ}|\kt\rangle_0
\eea
folded into the wave packet.  The last term has a single power of $\phi_0$ and so will not contribute to a term with no zero modes, and thus will play no role here.  Recall further that the terms in the parenthesis are associated with the meson wave packet, and so will be acted upon by the vacuum Hamiltonian, whereas $\vac$ is the dressed kink, which will be acted upon by the full kink Hamiltonian.   In other words, at second order our master formula (\ref{princ}) reads
\bea
0&=&(H\p_2-E_1)|\kt\rangle^\L_2\nonumber\\
&&-\frac{g V^{(3)}_-}{4\okt{}}\pin{k} \frac{\mb_{k} \mb_{\kt-k} \mb_{-\kt}}{\okt{}-\ok{}-\omega_{\kt-k}} [H^\L_3,\Bd{}B^\ddag_{\kt-k}]\vac_0+[H^\L_3,B^\ddag_{\kt}]\vac_1\nonumber\\
&&-\frac{g V^{(3)}_-}{4\okt{}}\pin{k} \frac{\mb_{k} \mb_{\kt-k} \mb_{-\kt}}{\okt{}-\ok{}-\omega_{\kt-k}} \Bd{}B^\ddag_{\kt-k}H\p_3\vac_0+B^\dag_{\kt}H\p_3\vac_1\nonumber\\
&&+[H^\L_4,B^\ddag_{\kt}]\vac_0+B^\ddag_{\kt}(H\p_4-E_2)\vac_0 \label{g2}
\eea
where $E_1=Q_1+\okt{}$.

Using the identity
\beq
[\phi^n(x),\Bd{}]=\frac{n\g_{-k}(x)}{2\ok{}}:\phi^{n-1}(x):_b
\eeq
one easily finds
\bea
[H^\L_3,B^\ddag_{\kt}]&=&\frac{g V^{(3)}_-}{4\okt{}}\int dx \g_{-\kt}(x)\left(:\phi^2(x):_b+\I(x)\right)\\
\left[H^{\L}_3 ,\Bd{}B^\ddag_{\kt-k}\right]
&=&\frac{g V^{(3)}_-}{4\ok{}}\int dx \g_{-k}(x)\left(:\phi^2(x):_b+\I(x)\right)B^\ddag_{\kt-k}\nonumber\\
&&+
\frac{g V^{(3)}_-}{4\omega_{\kt-k}}\int dx \g_{k-\kt}(x)\Bd{}\left(:\phi^2(x):_b+\I(x)\right).\nonumber
\eea
Similarly, using the $O(g^2)$ interaction
\beq
H^\L_4=\frac{g^2 V^{(4)}_-}{24}\int dx \left[:\phi^4(x):_b+6\I(x):\phi^2(x):_b+3\I^2(x)\right]
\eeq
one arrives at the commutator
\beq
[H^\L_4,B^\ddag_{\kt}]=\frac{g^2 V^{(4)}_-}{12\okt{}}\int dx \g_{-\kt}(x)\left[:\phi^3(x):_b+3\I(x)\phi(x)\right].
\eeq

\subsubsection{The First Term}

Now we are ready to evaluate the one-meson, no zero-mode contributions from all seven terms in the eigenvalue equation (\ref{g2}).  Using $E_1=Q_1+\okt{}$, the first term contains
\bea
(H\p_2-E_1)|\kt\rangle^\L_2
&\supset&\frac{\pi_0^2}{2}|\kt\rangle^{\L 21}_2
+\left(-\okt{}+\ppin{k} \ok{}\Bd{}B_{k}\right)|\kt\rangle^{\L 01}_2\label{c1}\\
&=&
\left(-\okt{}+\ppin{k} \ok{}\Bd{}B_{k}\right)|\kt\rangle^{\L 01}_2\nonumber\\
&&+\frac{1}{8Q_0}\left[4\kt^2+
\ppin{k}
\left(-|\Delta_{kB}|^2+g\sQ \frac{\Delta_{-kB} V_{\I k}}{\ok{}}
\right)\right.\nonumber\\
&&+\left.\frac{1}{2}\ppink{2}|\Delta_{k_1k_2}|^2 \frac{(\ok1-\ok2)^2}{\ok1\ok2}
\right]
|\kt\rangle_0.\nonumber
\eea
Note that the right hand side contains $|\kt\rangle^{\L 01}_2$, which is the second order coefficient that we are trying to find.  This is the only place that it will appear, as $|\kt\rangle^\L_2$ does not appear anywhere else in (\ref{g2}), and so we will not be able to fix the kernel of the term in the round parenthesis.  This kernel consists of the initial state $|\kt\rangle_0$ and also the state $|-\kt\rangle_0$, which is the coefficient corresponding to the leading quantum correction to elastic kink-meson scattering.  This ambiguity is physical, as the kink Hamiltonian indeed has two degenerate eigenstates and the two undetermined coefficients are just the weights of those states in $|\kt\rangle^\L$.  

In conclusion, we see that Eq.~(\ref{g2}) does not determine the $|-\kt\rangle_0$ component of $|\kt\rangle^\L_2$, reflecting an honest degeneracy in the spectrum of the Hamiltonian.  Rather, we brutally set this component to zero by hand using the condition that the wave packet moves rigidly whereas this component moves in the opposite direction.  So is this exercise trivial?  No, because the fact that a $|-\kt\rangle_0$ component of $|\kt\rangle^\L_2$ is in the kernel of the term in round parenthesis implies that the contribution from this term to the $|-\kt\rangle_0$ part of (\ref{g2}) cancels.  However, there may still be a contribution from the other six terms, and now we see that we cannot use $|\kt\rangle^\L_2$ to cancel it.  In other words, Eq.~(\ref{g2}) is overconstrained.  Therefore, it will be a nontrivial check of the consistency of (\ref{g2}), and therefore our master formula (\ref{princ}), that none of the other six terms include contributions proportional to $|-\kt\rangle_0$.

This is similar to the case of nonrelativistic quantum mechanics, where there are steady state solutions with any normalization and with any backwards traveling wave at $x=-\infty$.   As in that case, the freedom is just the freedom to choose an initial backwards scattering wave, and can be eliminated using the proper choice of boundary conditions.   If one solves the scattering problem using a kink Hamiltonian eigenstate, then the correct boundary condition is that there should be no incoming wave from the right.  If instead one solves the problem by evolving the incoming wave packet, then the correct boundary condition is that there should be no scattering before the wave packet arrives at the kink.  These two approaches to inelastic kink-meson scattering were described in Refs.~\cite{menorm} and \cite{memult} respectively where the corresponding boundary conditions were described.

\subsubsection{The Second Term}

Let us turn to the next term in Eq.~(\ref{g2}).  We need terms with one meson in the commutator $\left[H^{\L}_3 ,\Bd{}B^\ddag_{\kt-k}\right]$.  These include the zero-meson terms in $:\phi^2:_b$
\beq
:\phi^2(x):_b\supset \pink{2} \g_{k_1}(x)\g_{k_2}(x)\Bd{1}\frac{B_{-k_2}}{\ok{2}}.
\eeq
Therefore
\bea
\left[H^{\L}_3 ,\Bd{}B^\ddag_{\kt-k}\right]\vac_0&\supset&
\frac{g V^{(3)}_-}{4\ok{}}\int dx \g_{-k}(x)\left(\I(x)|\kt-k\rangle_0+\pin{k\p}\frac{ \g_{k\p}(x)\g_{k-\kt}(x)}{\omega_{k-\kt}}|k\p\rangle_0
\right)\nonumber\\
&&+
\frac{g V^{(3)}_-}{4\omega_{\kt-k}}\int dx \g_{k-\kt}(x)\I(x)|k\rangle_0.
\eea
The second term is therefore
\bea
-\frac{g V^{(3)}_-}{4\okt{}}\pin{k} \frac{\mb_{k} \mb_{\kt-k} \mb_{-\kt}}{\okt{}-\ok{}-\omega_{\kt-k}} [H^\L_3,\Bd{}B^\ddag_{\kt-k}]\vac_0&=&-\frac{g^2 V^{(3)2}_-}{16\okt{}}\pin{k} \frac{\mb_{k} \mb_{\kt-k} \mb_{-\kt}}{\ok{}(\okt{}-\ok{}-\omega_{\kt-k})} \nonumber\\
&&\hspace{-7cm}\times
\int dx \g_{-k}(x)\left(2\I(x)|\kt-k\rangle_0+\frac{ \g_{k-\kt}(x)}{\omega_{k-\kt}}\pin{k\p}\g_{k\p}(x)|k\p\rangle_0
\right). \label{t2}
\eea

This is the first term in which we have seen that the meson momentum is not manifestly conserved, because it is not proportional to $|\kt\rangle$.  The total momentum $P\p$ is conserved because $P\p$ commutes with $H\p$, which evolves the system.   Therefore, if the meson momentum is not conserved, it means that some momentum was exchanged with the kink.  However we do not expect this to happen if the meson is far from the kink.  To check this expectation, we should fold (\ref{t2}) into the meson wave packet.

Let us first look at the first term in the round parenthesis in Eq.~(\ref{t2}).  For $\sigma$ large enough, the $\omega$ terms may be evaluated at $k_0$ and removed from the $\kt$ integral, leaving a term proportional to
\beq
\int dx \I(x) \pin{\kt}e^{-i\kt x_0}\pin{k}  \g_k(x)|\kt-k\rangle_0=\int dx  \I(x) \pin{\kt}\pin{k}\g_k(x)e^{-i(\kt+k) x_0} |\kt\rangle_0.
\eeq
The phase of the $k$ integration varies quickly with respect to $k$, because $(d/dk)$Arg$(e^{-ikx_0})=-x_0$.  This variation is however canceled by that of the phase of $\g_k(x)$ if $x\sim x_0$.  Therefore, we learn that the $k$ integral is only appreciable when $x\sim x_0$.  However $\I(x_0)$ is exponentially suppressed in $mx_0$, so this term vanishes when folded into the wave packet in the large $mx_0$ limit.

A similar argument applies to the second term in Eq.~(\ref{t2}).  Now the $\kt$ integral is proportional to
\beq
\pin{\kt} e^{-i\kt x_0}\g_{k-\kt}(x)
\eeq
whose large $mx_0$ limit vanishes unless $x\sim x_0$.  In that case, one may replace all of the normal modes $\g(x)$ with their asymptotic forms (\ref{gk}).  Eq.~ (\ref{t2}) is then, up to terms annihilated by folding, equal to
\bea
&&-\frac{g^2 V^{(3)2}_-}{16\okt{}}\pin{k} \frac{  \mb_{-\kt}}{\ok{}\omega_{k-\kt}(\okt{}-\ok{}-\omega_{\kt-k})} 
\int dx{ e^{i\kt x}}{}\pin{k\p}\mb_{k\p}e^{-ik\p x}|k\p\rangle_0\nonumber\\
&&\hspace{2cm}=-\frac{g^2 V^{(3)2}_-}{16\okt{}}\left[\pin{k} \frac{  1}{\ok{}\omega_{k-\kt}(\okt{}-\ok{}-\omega_{\kt-k})} \right]
{}{}|\kt\rangle_0
.
\eea
This is proportional to $|\kt\rangle_0$, and so we see that far away from the kink, at $m x_0\gg 1$, the meson wave packet does not transfer momentum to the kink.  This of course is a property that we expect of an asymptotic incoming state, but not of a true kink Hamiltonian eigenstate, which will contain a scattered component.

\subsubsection{The Third Term}

The third term in the eigenvalue equation (\ref{g2}) is
\bea
[H^\L_3,B^\ddag_{\kt}]\vac_1
&\supset&\frac{g V^{(3)}_-}{4\okt{}}\int dx \g_{-\kt}(x)\\
&&\times\left[\I(x)\vac_1^{01}+\pink{2} \g_{k_1}(x)\g_{k_2}(x)\left(\frac{\Bd{1}B_{-k_2}}{\ok{2}}\vac_1^{01}+\frac{B_{-k_1}B_{-k_2}}{4\ok{1}\ok{2}}\vac_1^{03}\right)\right].\nonumber
\eea
Again, when folding into the wave packet, the $\kt$ integral vanishes in the $mx_0\rightarrow\infty$ limit unless $x\sim x_0$, in which case the $\I(x)$ term vanishes.

At large $x$  we use the asymptotic forms of the normal modes and perform the $x$ and $k_2$ integrations
\bea
[H^\L_3,B^\ddag_{\kt}]\vac_1
&\supset&\frac{g V^{(3)}_-}{4\okt{}} \mb_{-\kt}\pin{k} \mb_{k}\mb_{\kt-k}\left(\frac{\Bd{}B_{k-\kt}}{\omega_{k-\kt}}\vac_1^{01}+\frac{B_{-k}B_{k-\kt}}{4\ok{}\omega_{k-\kt}}\vac_1^{03}\right).\label{t3}
\eea
These terms involve interactions of the cloud around the kink, described by $\vac_1$, with the meson wave packet, described by the operators.  Needless to say, any such interaction would violate the locality that we require for our asymptotic states.  Therefore, it is an important consistency check of our choice of asymptotic state that these terms vanish when folded into the wave packet.

Let us look at the first term in the round parentheses, in which the dressed kink contains a single meson.  Folding it into the wave packet, one obtains
\bea
&&\sigma\sqrt{\pi}\frac{g^2 V^{(3)}_-}{2\okt{}} \pin{\kt}e^{-\sigma^2\left(\kt-k_0\right)^2}e^{i(k_0-\kt)x_0}\pin{k}\frac{\mb_{k+\kt}\mb_{-k}}{\ok{}} \left(\frac{V_{\I k}}{2\ok{}}+\frac{\Delta_{kB}}{2g\sQ}\right)|k+\kt\rangle_0\\
&&=\sigma\sqrt{\pi}\frac{g^2 V^{(3)}_-}{4\okt{}} \pin{k}\left[\pin{\kt}e^{-\sigma^2\left(\kt-k_0\right)^2}e^{i(k_0-\kt)x_0}\frac{\mb_{k}\mb_{\kt-k}}{\omega_{k-\kt}} \left(\frac{V_{\I, k-\kt}}{\omega_{k-\kt}}+\frac{\Delta_{k-\kt,B}}{g\sQ}\right)|k\rangle_0\right].\nonumber
\eea
The term $e^{-\sigma^2\left(\kt-k_0\right)^2}e^{i(k_0-\kt)x}$ varies very quickly with respect to $\kt$, whereas the other terms are essentially consistant within the support of the Gaussian $e^{-\sigma^2\left(\kt-k_0\right)^2}$ in the large $m\sigma$ limit.  Therefore we may replace $\kt$ with $k_0$ in the other terms and pull them out of the $\kt$ integral, leaving a Gaussian integral
\beq
\sigma\sqrt{\pi}\pin{\kt} e^{-\sigma^2\left(\kt-k_0\right)^2}e^{i(k_0-\kt)x_0}=\frac{e^{-x_0^2/(4\sigma^2)}}{2} \label{zero}
\eeq
which vanishes in the $x_0/\sigma\rightarrow\infty$ limit.  Thus, there is no nonlocal interaction between the kink and the wave packet arising from the $\vac_1^{01}$ term in (\ref{t3}).  In fact, like the $\I(x)$ term, it does not contribute to Eq.~(\ref{g2}).  

Finally we consider the $\vac_1^{03}$ term in (\ref{t3})
\bea
&&\frac{g^2 V^{(3)}_-}{24\okt{}} \mb_{-\kt}\pin{k} \mb_{k}\mb_{\kt-k}\frac{B_{-k}B_{k-\kt}}{4\ok{}\omega_{k-\kt}}
\ppink{3} \frac{V_{k_1k_2k_3}}{\ok 1+\ok 2+\ok 3}|k_1k_2k_3\rangle_0\\
&&\hspace{3cm}=\frac{g^2 V^{(3)}_-}{4\okt{}} \mb_{-\kt}\pin{k}\frac{ \mb_{k}\mb_{\kt-k}}{4\ok{}\omega_{k-\kt}}
\pin{k\p} \frac{V_{k\p,-k,k-\kt}}{\okp{}+\ok {}+\omega_{\kt-k}}|k\p\rangle_0.\nonumber
\eea
Again, everything varies slowly with respect to $\kt$ over the narrow support of the Gaussian $e^{-\sigma^2\left(\kt-k_0\right)^2}e^{i(k_0-\kt)x}$ in the wave packet $\alpha_\kt$, and so each $\kt$ here may be replaced by $k_0$ and the $\kt$ integral reduces to (\ref{zero}), vanishing in the $x_0/\sigma\rightarrow\infty$ limit.  

However, when $k\p\sim \kt$, there is a $\delta$-function divergence in $V_{k\p,-k,k-\kt}$
\beq
V_{k\p,-k,k-\kt}\supset V^{(3)}_-\int dx \mb_{k\p}\mb_{-k}\mb_{k-\kt}e^{i(\kt-k\p)x}=2\pi \delta(\kt-k\p)  V^{(3)}_- \mb_{k\p}\mb_{-k}\mb_{k-\kt}
\eeq
and so this contribution is not small.  It is
\beq
\frac{g^2 V^{(3) 2}_-}{16\okt{}} \pin{k}\frac{1}{\ok{}\omega_{k-\kt}(\okt{}+\ok {}+\omega_{\kt-k})} 
 |\kt\rangle_0.
\eeq


\subsubsection{The Fourth Term}

The fourth term contains $\Bd{}B^\ddag_{\kt-k}$ on the left.  As a result, it always leads to at least two mesons, so it cannot contribute to the one-meson correction to the asymptotic state.

\subsubsection{The Fifth Term}

Next we turn to the fifth term, which represents the second order corrected 
kink and the bare meson.  The terms with one meson and no zero modes, representing the energy correction to the kink, are
\bea
B^\dag_{\kt}H\p_3\vac_1
&=&B^\dag_{\kt} \left[
\frac{g}{6}\ppink{3} V_{k_1k_2k_3}\frac{B_{-k_1}B_{-k_2}B_{-k_3}}{8\ok{1}\ok{2}\ok{3}}+\frac{g}{2}\ppin{k} V_{\I k}\frac{B_{-k}}{2\ok{}}
\right]\\
&&\times
\left[
-\frac{g}{6}\ppinkp{3} \frac{V_{k\p_1k\p_2k\p_3}}{\okp 1+\okp 2+\okp 3}|k\p_1k\p_2k\p_3\rangle_0-\frac{g}{2}\ppin{k\p} \left(\frac{V_{\I k\p}}{\okp{}}+\frac{\Delta_{k\p B}}{g\sQ}\right)|k\p\rangle_0
\right]\nonumber\\
&&\hspace{-2cm}=\left[
-\frac{g^2}{48}\ppink{3}\frac{|V_{k_1k_2k_3}|^2}{\ok1\ok2\ok3(\ok1+\ok2+\ok3)}
-\frac{g^2}{8}\ppin{k}\left(|V_{\I k}|^2+\frac{V_{\I k}\Delta_{-k B}}{g\sQ}
\right)
\right]|\kt\rangle_0.\nonumber
\eea
Note that the $V\Delta$ cross-term cancels that in the first contribution (\ref{c1}).

\subsubsection{The Sixth Term}

The sixth term represents the second order correction to the meson together with the bare kink.  The one-meson part is
\bea
[H^\L_4,B^\ddag_{\kt}]\vac_0
&\supset&\frac{g^2 V^{(4)}_-}{4\okt{}}\int dx \g_{-\kt}(x)\I(x)\phi(x)\vac_0\\
&=&\frac{g^2 V^{(4)}_-}{4\okt{}}\int dx \g_{-\kt}(x)\I(x)\ppin{k}\g_k(x)|k\rangle_0.\nonumber
\eea
As usual, when folded into the wave packet, the $\kt$ integration implies that this is nonvanishing only for $x\sim x_0$ and so $\I(x)\rightarrow 0$ for a well-separated initial meson wave packet.  This is to be expected, as $\I(x)$ results from the interaction of the meson with the kink, which is too far to interact.  Thus there is no such contribution, the momentum of an isolated meson is conserved far from the kink.

\subsubsection{The Seventh Term}

The last term represents the second order correction to the kink with a bare meson, together with the correction to the energy.  The only terms contributing a single meson are
\bea
B^\ddag_{\kt}(H\p_4-E_2)\vac_0=\left(\frac{g^2V_{\I\I}}{8}-E_2\right)|\kt\rangle_0. \label{t7}
\eea

\subsection{Defining $E$}

We have seen that the only terms in Eq.~(\ref{g2}) which survive are proportional to $|\kt\rangle_0$.  This means that the momentum of our asymptotic state is conserved when it is far from the kink, as one expects.  Recall that the second order contribution $|\kt\rangle^{\L 01}_2$ does not contribute to $|\kt\rangle_0$ terms in (\ref{g2}), as they are annihilated by the free kinetic term minus $\okt{}$.  Therefore, the only unknown in (\ref{g2}) is $E_2$ in the seventh contribution (\ref{t7}), and fixing $E_2$ to the sum of the other terms will lead (\ref{g2}) to be satisfied.

This is easily done.  One finds
\beq
E_2=Q_2+M_2
\eeq
where
\bea
Q_2&=&\frac{g^2V_{\I\I}}{8}-\frac{g^2}{48}\ppink{3}\frac{|V_{k_1k_2k_3}|^2}{\ok1\ok2\ok3(\ok1+\ok2+\ok3)}\\
&&
-\frac{g^2}{8}\ppin{k}|V_{\I k}|^2-\frac{1}{8Q_0}\ppin{k}|\Delta_{kB}|^2+\frac{1}{16}\ppink{2}|\Delta_{k_1k_2}|^2 \frac{(\ok1-\ok2)^2}{\ok1\ok2}
\nonumber
\eea
is the 2-loop correction to the ground state kink mass, first found in Ref.~\cite{me2loop}.  On the other hand
\beq
M_2=-\frac{g^2 V^{(3)2}_-}{8\okt{}}\left[\pin{k} \frac{ \ok{}+\omega_{\kt-k}}{\ok{}\omega_{k-\kt}((\ok{}+\omega_{\kt-k})^2-\omega^2_\kt)} \right]
{}{}
\eeq
is the one-loop correction to the moving meson energy in the left vacuum.  

The variable $E_2$ is part of the definition of our asymptotic state $|\kt\rangle^\L$, and so this is an essential result.  We believe that more generally the variable $E$ can be written as the sum of the exact quantum corrections to the kink mass, calculated using the kink Hamiltonian, with the corrections to the energy of a moving meson that one would calculate in the vacuum sector where the meson wave packet is located.  

In summary, we have found $E_2$ and we have shown that $|\kt\rangle_2^{\L 01}$ is proportional to $|\kt\rangle_0$, with an arbitrary coefficient reflecting a choice of normalization of $|\kt\rangle^\L$.  There is no $|-\kt\rangle_0$ piece, and so no corresponding contribution to elastic meson-kink scattering.

\section{Remarks}

Our goal in this work has been to define asymptotic states consisting of a kink and a meson wave packet far to its left which have two properties.  First,  the state evolves via a constant velocity motion of the meson wave packet, keeping its shape including all quantum corrections, up to the usual wave packet spreading effects.   Second, the state is invariant under rigid translations, which translate both the kink and the meson while maintaining their relative distance. 

We then presented a construction.  Translation-invariance was manifest in our construction.  It fixed a part of our state, which we called the descendant.  The rest of the state, called the primary, was fixed via a variant on the usual eigenvalue problem.  Instead of imposing that the state be an eigenstate of the full Hamiltonian, we separated it into a dressed kink and a dressed meson and acted on the dressed meson with a vacuum Hamiltonian.  This is a truncation of the full Hamiltonian corresponding, intuitively, to removing the kink.  

This is not clearly consistent, as it is inevitable that the vacuum Hamiltonian terms also contract with the dressed kink and also evolution proceeds via the full Hamiltonian.  Also, it is not obvious that this leads to constant velocity, rigid evolution.  However, we showed that at the first few orders the potentially offending terms disappear when folded into the wave packet.  Thus, our construction appears to satisfy not only the translation-invariance criterion but also the rigid motion criterion, as desired.

The generalization to an arbitrary number of meson wave packets in arbitrary positions is obvious.  One simply keeps track of which wave packet which virtual meson is associated with, and uses the vacuum Hamiltonian corresponding to the vacuum in its position.  Similarly, we believe that the generalization to extended domain walls is straightforward.  We hope that in the future this approach may be generalized to the scattering of general global solitons, such as Skyrmions, with elementary quanta, as this will teach us about baryon-meson scattering~\cite{hayashi1}.

The results here justify some assumptions made in Refs.~\cite{memult,menorm,mestokes} regarding quantum corrections to the initial and final states in inelastic meson-kink scattering.  In the short term, this was the last missing ingredient needed for the study of several problems.  These problems include higher order corrections to inelastic meson-kink scattering, leading order elastic meson-kink scattering and a determination of quantum corrections to the life time of an overly excited shape mode, for example the twice-excited shape mode in the $\phi^4$ model.  Such lifetimes are in general dependent on the quantum corrections to the unstable state, but can be made well-defined as the widths of resonances in elastic meson-kink scattering.  We hope to turn to the study of such unstable resonances in the near future.

One may ask whether our asymptotic states may be created directly from $H\p$ eigenstates, avoiding our construction.  Of course these are a basis of states, so some such construction must be possible.  Nonetheless, we have seen that the use of $H^\L$ here removes on-shell, degenerate eigenstates which, using $H\p$, needed to be removed by hand during a careful matching in Ref.~\cite{menorm}.  Also, $H^\L$ leads to much simpler expressions, such as (\ref{init}), for states, manifestly eliminating $\Delta_{Bk}$ terms and shape modes from the operator creating the dressed wave packet and that this operator, at each order and in each term, contains a product of $\Bd{i}$ operators such that $\sum_i k_i=\kt$.  Furthermore, our asymptotic states can naturally be tensored together to create any number of mesons on both sides of the kink, whereas states constructed from $H\p$ will necessarily have on-shell additional mesons on one side of the kink or the other, with the numbers on each side differing by the number created during scattering.

Needless to say, with these asymptotic states in hand, it would be tempting to search for an LSZ reduction formula valid in the one-kink sector, which allows the states $|\kt\rangle^\L$ to be replaced by $|\kt\rangle_0$ while removing those irreducible interactions which occur far from and on the same side of the kink.

\section* {Acknowledgement}

\noindent
JE is supported by NSFC MianShang grants 11875296 and 11675223.


\begin{thebibliography}{99}

\bibitem{csw} D.~K. Campbell, J.~F. Schonfeld and C.~A. Wingate,
``Resonance structure in kink-antikink interactions in $\phi^4$ theory,"
Physica \textbf{D9} (1983) 1.


\bibitem{doreyf6}
P.~Dorey, K.~Mersh, T.~Romanczukiewicz and Y.~Shnir,
``Kink-antikink collisions in the $\phi^6$ model,''
Phys. Rev. Lett. \textbf{107} (2011), 091602
doi:10.1103/PhysRevLett.107.091602
[arXiv:1101.5951 [hep-th]].

\bibitem{sfal21}
C.~Adam, D.~Ciurla, K.~Oles, T.~Romanczukiewicz and A.~Wereszczynski,
``Sphalerons and resonance phenomenon in kink-antikink collisions,''
Phys. Rev. D \textbf{104} (2021) no.10, 105022
doi:10.1103/PhysRevD.104.105022
[arXiv:2109.01834 [hep-th]].

\bibitem{multex22a}
F.~C.~Simas, K.~Z.~Nobrega, D.~Bazeia and A.~R.~Gomes,
``Degeneracy and kink scattering in a two coupled scalar field model in $(1,1)$ dimensions,''
[arXiv:2201.03372 [hep-th]].

\bibitem{multex22b}
A.~Moradi Marjaneh, F.~C.~Simas and D.~Bazeia,
``Collisions of kinks in deformed \ensuremath{\varphi}4 and \ensuremath{\varphi}6 models,''
``Collisions of kinks in deformed  \ensuremath{\varphi}${}^4$ and \ensuremath{\varphi}${}^6$ models,''
Chaos Solitons and Fractals: the interdisciplinary journal of Nonlinear Science and Nonequilibrium and Complex Phenomena \textbf{164} (2022), 112723
doi:10.1016/j.chaos.2022.112723
[arXiv:2207.00835 [hep-th]].

\bibitem{alonso22}
A.~Alonso-Izquierdo, D.~Migu\'elez-Caballero, L.~M.~Nieto and J.~Queiroga-Nunes,
``Wobbling kinks in a two-component scalar field theory: Interaction between shape modes,''
Physica D: Nonlinear Phenomena \textbf{443} (2023), 133590
doi:10.1016/j.physd.2022.133590
[arXiv:2207.10989 [hep-th]].



\bibitem{col22}
C.~Adam, P.~Dorey, A.~Garcia Martin-Caro, M.~Huidobro, K.~Oles, T.~Romanczukiewicz, Y.~Shnir and A.~Wereszczynski,
``Multikink scattering in the $\phi^6$ model revisited,''
[arXiv:2209.08849 [hep-th]].

\bibitem{tanmay21}
M.~Mukhopadhyay, E.~I.~Sfakianakis, T.~Vachaspati and G.~Zahariade,
``Kink-antikink scattering in a quantum vacuum,''
JHEP \textbf{04} (2022), 118
doi:10.1007/JHEP04(2022)118
[arXiv:2110.08277 [hep-th]].

\bibitem{takyi22}
I.~Takyi and H.~Weigel,
``Quantum effects of solitons in the self-dual impurity model,''
Phys. Rev. D \textbf{107} (2023) no.3, 036003
doi:10.1103/PhysRevD.107.036003
[arXiv:2212.02332 [hep-th]].

\bibitem{tanmay23}
M.~Mukhopadhyay and T.~Vachaspati,
``Resonance structures in kink-antikink scattering in a quantum vacuum,''
[arXiv:2303.03415 [hep-th]].

\bibitem{tomrad1}
T.~Romanczukiewicz,
``Interaction between kink and radiation in phi**4 model,''
Acta Phys. Polon. B \textbf{35} (2004), 523-540
[arXiv:hep-th/0303058 [hep-th]].

\bibitem{tomrad2}
T.~Romanczukiewicz,
``Interaction between topological defects and radiation,''
Acta Phys. Polon. B \textbf{36} (2005), 3877-3887

\bibitem{tomrad3}
P.~Forgacs, A.~Lukacs and T.~Romanczukiewicz,
``Negative radiation pressure exerted on kinks,''
Phys. Rev. D \textbf{77} (2008), 125012
doi:10.1103/PhysRevD.77.125012
[arXiv:0802.0080 [hep-th]].

\bibitem{faddeev77}
L.~D.~Faddeev and V.~E.~Korepin,
``Quantum Theory of Solitons: Preliminary Version,''
Phys. Rept. \textbf{42} (1978), 1-87
doi:10.1016/0370-1573(78)90058-3

\bibitem{lowe79}
M.~Lowe,
``BOSON - SOLITON SCATTERING IN THE SINE-GORDON MODEL,''
Nucl. Phys. B \textbf{159} (1979), 349-362
doi:10.1016/0550-3213(79)90339-0

\bibitem{parm87}
J.~A.~Parmentola and I.~Zahed,
``MESON - SOLITON SCATTERING WITH SOLITON RECOIL,''
Print-87-0301 (STONY BROOK).

\bibitem{swanson88}
M.~S.~Swanson,
``SOLITON-PARTICLE SCATTERING AND BERRY'S PHASE,''
Phys. Rev. D \textbf{38} (1988), 3122-3127
doi:10.1103/PhysRevD.38.3122

\bibitem{uehara91}
M.~Uehara, A.~Hayashi and S.~Saito,
``Meson - soliton scattering with full recoil in standard collective coordinate quantization,''
Nucl. Phys. A \textbf{534} (1991), 680-696
doi:10.1016/0375-9474(91)90466-J


\bibitem{hayashi1}
A.~Hayashi, S.~Saito and M.~Uehara,
``Pion - nucleon scattering in the Skyrme model and the P wave Born amplitudes,''
Phys. Rev. D \textbf{43} (1991), 1520-1531
doi:10.1103/PhysRevD.43.1520

\bibitem{hayashi2}
A.~Hayashi, S.~Saito and M.~Uehara,
``Pion - nucleon scattering in the soliton model,''
Prog. Theor. Phys. Suppl. \textbf{109} (1992), 45-72
doi:10.1143/PTPS.109.45


\bibitem{abdel11}
A.~M.~H.~H.~Abdelhady and H.~Weigel,
``Wave-Packet Scattering off the Kink-Solution,''
Int. J. Mod. Phys. A \textbf{26} (2011), 3625-3640
doi:10.1142/S0217751X11054012
[arXiv:1106.3497 [nlin.PS]].

\bibitem{merefl}
J.~Evslin and H.~Liu,
``Quantum Reflective Kinks,''
[arXiv:2210.12725 [hep-th]].

\bibitem{memult}
H.~Liu, J.~Evslin and B.~Zhang,
``Meson production from kink-meson scattering,''
Phys. Rev. D \textbf{107} (2023) no.2, 025012
doi:10.1103/PhysRevD.107.025012
[arXiv:2211.01794 [hep-th]].

\bibitem{mestokes}
J.~Evslin and H.~Liu,
``(Anti-)Stokes scattering on kinks,''
JHEP \textbf{03} (2023), 095
doi:10.1007/JHEP03(2023)095
[arXiv:2301.04099 [hep-th]].

\bibitem{menorm}
J.~Evslin and H.~Liu,
``A reduced inner product for kink states,''
JHEP \textbf{03} (2023), 070
doi:10.1007/JHEP03(2023)070
[arXiv:2212.10344 [hep-th]].

\bibitem{skyrme}
T.~H.~R.~Skyrme,
``A Nonlinear field theory,''
Proc. Roy. Soc. Lond. A \textbf{260} (1961), 127-138
doi:10.1098/rspa.1961.0018

\bibitem{smorg}
S.~B.~Gudnason and C.~Halcrow,
``A Sm\"orgasbord of Skyrmions,''
JHEP \textbf{08} (2022), 117
doi:10.1007/JHEP08(2022)117
[arXiv:2202.01792 [hep-th]].

\bibitem{mekink}
J.~Evslin,
``Manifestly Finite Derivation of the Quantum Kink Mass,''
JHEP \textbf{11} (2019), 161
doi:10.1007/JHEP11(2019)161
[arXiv:1908.06710 [hep-th]].

\bibitem{me2loop}
J.~Evslin and H.~Guo,
``Two-Loop Scalar Kinks,''
Phys. Rev. D \textbf{103} (2021) no.12, 125011
doi:10.1103/PhysRevD.103.125011
[arXiv:2012.04912 [hep-th]].

\bibitem{wstabile}
H.~Weigel,
``Quantum Instabilities of Solitons,''
AIP Conf. Proc. \textbf{2116} (2019) no.1, 170002
doi:10.1063/1.5114153
[arXiv:1907.10942 [hep-th]].

\bibitem{cahill76}
K.~E.~Cahill, A.~Comtet and R.~J.~Glauber,
``Mass Formulas for Static Solitons,''
Phys. Lett. B \textbf{64} (1976), 283-285
doi:10.1016/0370-2693(76)90202-1

\bibitem{cl75}
N.~H.~Christ and T.~D.~Lee,
``Quantum Expansion of Soliton Solutions,''
Phys. Rev. D \textbf{12} (1975), 1606
doi:10.1103/PhysRevD.12.1606

\bibitem{gjscc}
J.~L.~Gervais, A.~Jevicki and B.~Sakita,
``Collective Coordinate Method for Quantization of Extended Systems,''
Phys. Rept. \textbf{23} (1976), 281-293
doi:10.1016/0370-1573(76)90049-1


\bibitem{mewick}
J.~Evslin,
``Normal ordering normal modes,''
Eur. Phys. J. C \textbf{81} (2021) no.1, 92
doi:10.1140/epjc/s10052-021-08890-7
[arXiv:2007.05741 [hep-th]].

\bibitem{rebhan}
  A.~Rebhan and P.~van Nieuwenhuizen,
  ``No saturation of the quantum Bogomolnyi bound by two-dimensional supersymmetric solitons,''
  Nucl.\ Phys.\ B {\bf 508} (1997) 449
  doi:10.1016/S0550-3213(97)00625-1, 10.1016/S0550-3213(97)80021-1
 [hep-th/9707163].

\bibitem{gj76}
J.~L.~Gervais and A.~Jevicki,
``Point Canonical Transformations in Path Integral,''
Nucl. Phys. B \textbf{110} (1976), 93-112
doi:10.1016/0550-3213(76)90422-3

\end{thebibliography}

\begin{thebibliography}{99}

\bibitem{gjscc}
J.~L.~Gervais, A.~Jevicki and B.~Sakita,
``Collective Coordinate Method for Quantization of Extended Systems,''
Phys. Rept. \textbf{23} (1976), 281-293
doi:10.1016/0370-1573(76)90049-1

\bibitem{dhn2}
  R.~F.~Dashen, B.~Hasslacher and A.~Neveu,
  ``Nonperturbative Methods and Extended Hadron Models in Field Theory 2. Two-Dimensional Models and Extended Hadrons,''
  Phys.\ Rev.\ D {\bf 10} (1974) 4130.
 doi:10.1103/PhysRevD.10.4130

\bibitem{wrev}
N.~Graham and H.~Weigel,
``Quantum corrections to soliton energies,''
Int. J. Mod. Phys. A \textbf{37} (2022) no.19, 2241004
doi:10.1142/S0217751X22410044
[arXiv:2201.12131 [hep-th]].

\bibitem{gj76}
J.~L.~Gervais and A.~Jevicki,
``Point Canonical Transformations in Path Integral,''
Nucl. Phys. B \textbf{110} (1976), 93-112
doi:10.1016/0550-3213(76)90422-3

\bibitem{vega}
H.~J.~de Vega,
``Two-Loop Quantum Corrections to the Soliton Mass in Two-Dimensional Scalar Field Theories,''
Nucl. Phys. B \textbf{115} (1976), 411-428
doi:10.1016/0550-3213(76)90497-1

\bibitem{verwaest}
J.~Verwaest,
``Higher Order Correction to the Sine-Gordon Soliton Mass,''
Nucl. Phys. B \textbf{123} (1977), 100-108
doi:10.1016/0550-3213(77)90343-1

\bibitem{shifmananomalia}
M.~A.~Shifman, A.~I.~Vainshtein and M.~B.~Voloshin,
``Anomaly and quantum corrections to solitons in two-dimensional theories with minimal supersymmetry,''
Phys. Rev. D \textbf{59} (1999), 045016
doi:10.1103/PhysRevD.59.045016
[arXiv:hep-th/9810068 [hep-th]].

\bibitem{hayashi1}
A.~Hayashi, S.~Saito and M.~Uehara,
``Pion - nucleon scattering in the Skyrme model and the P wave Born amplitudes,''
Phys. Rev. D \textbf{43} (1991), 1520-1531
doi:10.1103/PhysRevD.43.1520

\bibitem{hayashi2}
A.~Hayashi, S.~Saito and M.~Uehara,
``Pion - nucleon scattering in the soliton model,''
Prog. Theor. Phys. Suppl. \textbf{109} (1992), 45-72
doi:10.1143/PTPS.109.45


\bibitem{accel}
I.~V.~Melnikov, C.~Papageorgakis and A.~B.~Royston,
``Accelerating solitons,''
Phys. Rev. D \textbf{102} (2020) no.12, 125002
doi:10.1103/PhysRevD.102.125002
[arXiv:2007.11028 [hep-th]].

\bibitem{andyprl}
I.~V.~Melnikov, C.~Papageorgakis and A.~B.~Royston,
``Forced Soliton Equation and Semiclassical Soliton Form Factors,''
Phys. Rev. Lett. \textbf{125} (2020) no.23, 231601
doi:10.1103/PhysRevLett.125.231601
[arXiv:2010.10381 [hep-th]].

\bibitem{raggio}
J.~F.~Wheater and P.~D.~Xavier,
``The Size of a Soliton,''
[arXiv:2207.01274 [hep-th]].

\bibitem{mekink}
J.~Evslin,
``Manifestly Finite Derivation of the Quantum Kink Mass,''
JHEP \textbf{11} (2019), 161
doi:10.1007/JHEP11(2019)161
[arXiv:1908.06710 [hep-th]].

\bibitem{me2loop}
J.~Evslin and H.~Guo,
``Two-Loop Scalar Kinks,''
Phys. Rev. D \textbf{103} (2021) no.12, 125011
doi:10.1103/PhysRevD.103.125011
[arXiv:2012.04912 [hep-th]].


\bibitem{memono}
J.~Evslin and S.~B.~Gudnason,
``Dwarf Galaxy Sized Monopoles as Dark Matter?,''
[arXiv:1202.0560 [astro-ph.CO]].


\bibitem{fuzzy14}
H.~Y.~Schive, T.~Chiueh and T.~Broadhurst,
``Cosmic Structure as the Quantum Interference of a Coherent Dark Wave,''
Nature Phys. \textbf{10} (2014), 496-499
doi:10.1038/nphys2996
[arXiv:1406.6586 [astro-ph.GA]].

\bibitem{impure}
C.~Adam, K.~Oles, J.~M.~Queiruga, T.~Romanczukiewicz and A.~Wereszczynski,
``Solvable self-dual impurity models,''
JHEP \textbf{07} (2019), 150
doi:10.1007/JHEP07(2019)150
[arXiv:1905.06080 [hep-th]].

\bibitem{chris}
J.~Evslin, C.~Halcrow, T.~Romanczukiewicz and A.~Wereszczynski,
``Spectral walls at one loop,''
Phys. Rev. D \textbf{105} (2022) no.12, 125002
doi:10.1103/PhysRevD.105.125002
[arXiv:2202.08249 [hep-th]].

\bibitem{hayashirep}
B.~Schwesinger, H.~Weigel, G.~Holzwarth and A.~Hayashi,
``The Skyrme Soliton in Pion, Vector and Scalar Meson Fields: $\pi N$ Scattering and Photoproduction,''
Phys. Rept. \textbf{173} (1989), 173
doi:10.1016/0370-1573(89)90022-7

\bibitem{skyrme}
T.~H.~R.~Skyrme,
``A Nonlinear field theory,''
Proc. Roy. Soc. Lond. A \textbf{260} (1961), 127-138
doi:10.1098/rspa.1961.0018

\bibitem{smorg}
S.~B.~Gudnason and C.~Halcrow,
``A Sm\"orgasbord of Skyrmions,''
JHEP \textbf{08} (2022), 117
doi:10.1007/JHEP08(2022)117
[arXiv:2202.01792 [hep-th]].

\bibitem{point22}
M.~A.~A.~Martin, R.~Schlesier and J.~Zahn,
``The semiclassical energy density of kinks and solitons,''
[arXiv:2204.08785 [hep-th]].

\bibitem{phi42loop}
J.~Evslin,
``\ensuremath{\phi^4} kink mass at two loops,''
Phys. Rev. D \textbf{104} (2021) no.8, 085013
doi:10.1103/PhysRevD.104.085013
[arXiv:2104.07991 [hep-th]].

\bibitem{menormal}
J.~Evslin and H.~Guo,
``Excited Kinks as Quantum States,''
Eur. Phys. J. C \textbf{81} (2021) no.10, 936
doi:10.1140/epjc/s10052-021-09739-9
[arXiv:2104.03612 [hep-th]].

\bibitem{hengyuan}
H.~Guo,
``Leading quantum correction to the \ensuremath{\Phi}4 kink form factor,''
Phys. Rev. D \textbf{106} (2022) no.9, 096001
doi:10.1103/PhysRevD.106.096001
[arXiv:2209.03650 [hep-th]].

\bibitem{alberto}
J.~Evslin and A.~Garc\'\i{}a Mart\'\i{}n-Caro,
``Spontaneous emission from excited quantum kinks,''
JHEP \textbf{12} (2022), 111
doi:10.1007/JHEP12(2022)111
[arXiv:2210.13791 [hep-th]].

\bibitem{memult}
H.~Liu, J.~Evslin and B.~Zhang,
``Meson production from kink-meson scattering,''
Phys. Rev. D \textbf{107} (2023) no.2, 025012
doi:10.1103/PhysRevD.107.025012
[arXiv:2211.01794 [hep-th]].


\bibitem{wstabile}
H.~Weigel,
``Quantum Instabilities of Solitons,''
AIP Conf. Proc. \textbf{2116} (2019) no.1, 170002
doi:10.1063/1.5114153
[arXiv:1907.10942 [hep-th]].

\bibitem{cahill76}
K.~E.~Cahill, A.~Comtet and R.~J.~Glauber,
``Mass Formulas for Static Solitons,''
Phys. Lett. B \textbf{64} (1976), 283-285
doi:10.1016/0370-2693(76)90202-1

\bibitem{mewick}
J.~Evslin,
``Normal ordering normal modes,''
Eur. Phys. J. C \textbf{81} (2021) no.1, 92
doi:10.1140/epjc/s10052-021-08890-7
[arXiv:2007.05741 [hep-th]].

\bibitem{shapeinter}
A.~Alonso-Izquierdo, D.~Migu\'elez-Caballero, L.~M.~Nieto and J.~Queiroga-Nunes,
``Wobbling kinks in a two-component scalar field theory: Interaction between shape modes,''
[arXiv:2207.10989 [hep-th]].

\bibitem{wshifman}
H.~Weigel and N.~Graham,
``Vacuum polarization energy of the Shifman\textendash{}Voloshin soliton,''
Phys. Lett. B \textbf{783} (2018), 434-439
doi:10.1016/j.physletb.2018.07.027
[arXiv:1806.07584 [hep-th]].

\bibitem{takyi1}
I.~Takyi, M.~K.~Matfunjwa and H.~Weigel,
``Quantum corrections to solitons in the $\Phi^8$ model,''
Phys. Rev. D \textbf{102} (2020) no.11, 116004
doi:10.1103/PhysRevD.102.116004
[arXiv:2010.07182 [hep-th]].

\bibitem{takyi2}
I.~Takyi, B.~Barnes and J.~Ackora-Prah,
``Vacuum Polarization Energy of the Kinks in the Sinh-Deformed Models,''
Turk. J. Phys. \textbf{45} (2021), 194-206
[arXiv:2012.12343 [hep-th]].

\bibitem{yuan1}
Y.~Zhong,
``Normal modes for two-dimensional gravitating kinks,''
Phys. Lett. B \textbf{827} (2022), 136947
doi:10.1016/j.physletb.2022.136947
[arXiv:2112.08683 [hep-th]].

\bibitem{yuan2}
Y.~Zhong,
``Singular P\"oschl-Teller II potentials and gravitating kinks,''
JHEP \textbf{09} (2022), 165
doi:10.1007/JHEP09(2022)165
[arXiv:2207.12681 [hep-th]].

\bibitem{quantosc}
M.~P.~Hertzberg,
``Quantum Radiation of Oscillons,''
Phys. Rev. D \textbf{82} (2010), 045022
doi:10.1103/PhysRevD.82.045022
[arXiv:1003.3459 [hep-th]].

\bibitem{kovbreather}
A.~Kovtun,
``Analytical computation of quantum corrections to a nontopological soliton within the saddle-point approximation,''
Phys. Rev. D \textbf{105} (2022) no.3, 036011
doi:10.1103/PhysRevD.105.036011
[arXiv:2110.05222 [hep-th]].

\end{thebibliography}

\begin{thebibliography}{99}

\bibitem{csw} D.~K. Campbell, J.~F. Schonfeld and C.~A. Wingate,
``Resonance structure in kink-antikink interactions in $\phi^4$ theory,"
Physica \textbf{D9} (1983) 1.

\bibitem{osc}
H.~Segur and M.~D.~Kruskal,
``Nonexistence of Small Amplitude Breather Solutions in $\phi^4$ Theory,''
Phys. Rev. Lett. \textbf{58} (1987), 747-750
doi:10.1103/PhysRevLett.58.747

\bibitem{osc3d}
G.~Fodor, P.~Forgacs, P.~Grandclement and I.~Racz,
``Oscillons and Quasi-breathers in the phi**4 Klein-Gordon model,''
Phys. Rev. D \textbf{74} (2006), 124003
doi:10.1103/PhysRevD.74.124003
[arXiv:hep-th/0609023 [hep-th]].

\bibitem{doreyf6}
P.~Dorey, K.~Mersh, T.~Romanczukiewicz and Y.~Shnir,
``Kink-antikink collisions in the $\phi^6$ model,''
Phys. Rev. Lett. \textbf{107} (2011), 091602
doi:10.1103/PhysRevLett.107.091602
[arXiv:1101.5951 [hep-th]].

\bibitem{multex22a}
F.~C.~Simas, K.~Z.~Nobrega, D.~Bazeia and A.~R.~Gomes,
``Degeneracy and kink scattering in a two coupled scalar field model in $(1,1)$ dimensions,''
[arXiv:2201.03372 [hep-th]].

\bibitem{multex22b}
A.~Moradi Marjaneh, F.~C.~Simas and D.~Bazeia,
``Collisions of kinks in deformed  \ensuremath{\varphi}${}^4$ and \ensuremath{\varphi}${}^6$ models,''
Chaos Solitons and Fractals: the interdisciplinary journal of Nonlinear Science and Nonequilibrium and Complex Phenomena \textbf{164} (2022), 112723
doi:10.1016/j.chaos.2022.112723
[arXiv:2207.00835 [hep-th]].

\bibitem{sfal21}
C.~Adam, D.~Ciurla, K.~Oles, T.~Romanczukiewicz and A.~Wereszczynski,
``Sphalerons and resonance phenomenon in kink-antikink collisions,''
Phys. Rev. D \textbf{104} (2021) no.10, 105022
doi:10.1103/PhysRevD.104.105022
[arXiv:2109.01834 [hep-th]].

\bibitem{col22}
C.~Adam, P.~Dorey, A.~Garcia Martin-Caro, M.~Huidobro, K.~Oles, T.~Romanczukiewicz, Y.~Shnir and A.~Wereszczynski,
``Multikink scattering in the $\phi^6$ model revisited,''
[arXiv:2209.08849 [hep-th]].

\bibitem{mm}
N.~S.~Manton and H.~Merabet,
``$\phi^4$ kinks: Gradient flow and dynamics,''
Nonlinearity \textbf{10} (1997), 3
doi:10.1088/0951-7715/10/1/002
[arXiv:hep-th/9605038 [hep-th]].

\bibitem{tomrad1}
T.~Romanczukiewicz,
``Interaction between kink and radiation in phi**4 model,''
Acta Phys. Polon. B \textbf{35} (2004), 523-540
[arXiv:hep-th/0303058 [hep-th]].

\bibitem{tomrad2}
T.~Romanczukiewicz,
``Interaction between topological defects and radiation,''
Acta Phys. Polon. B \textbf{36} (2005), 3877-3887

\bibitem{tomrad3}
P.~Forgacs, A.~Lukacs and T.~Romanczukiewicz,
``Negative radiation pressure exerted on kinks,''
Phys. Rev. D \textbf{77} (2008), 125012
doi:10.1103/PhysRevD.77.125012
[arXiv:0802.0080 [hep-th]].

\bibitem{mekink}
J.~Evslin,
``Manifestly Finite Derivation of the Quantum Kink Mass,''
JHEP \textbf{11} (2019), 161
doi:10.1007/JHEP11(2019)161
[arXiv:1908.06710 [hep-th]].

\bibitem{me2loop}
J.~Evslin and H.~Guo,
``Two-Loop Scalar Kinks,''
Phys. Rev. D \textbf{103} (2021) no.12, 125011
doi:10.1103/PhysRevD.103.125011
[arXiv:2012.04912 [hep-th]].

\bibitem{gjscc}
J.~L.~Gervais, A.~Jevicki and B.~Sakita,
``Collective Coordinate Method for Quantization of Extended Systems,''
Phys. Rept. \textbf{23} (1976), 281-293
doi:10.1016/0370-1573(76)90049-1

\bibitem{gj76}
J.~L.~Gervais and A.~Jevicki,
``Point Canonical Transformations in Path Integral,''
Nucl. Phys. B \textbf{110} (1976), 93-112
doi:10.1016/0550-3213(76)90422-3

\bibitem{wstabile}
H.~Weigel,
``Quantum Instabilities of Solitons,''
AIP Conf. Proc. \textbf{2116} (2019) no.1, 170002
doi:10.1063/1.5114153
[arXiv:1907.10942 [hep-th]].

\bibitem{cahill76}
K.~E.~Cahill, A.~Comtet and R.~J.~Glauber,
``Mass Formulas for Static Solitons,''
Phys. Lett. B \textbf{64} (1976), 283-285
doi:10.1016/0370-2693(76)90202-1

\bibitem{phi42loop}
J.~Evslin,
``\ensuremath{\phi^4} kink mass at two loops,''
Phys. Rev. D \textbf{104} (2021) no.8, 085013
doi:10.1103/PhysRevD.104.085013
[arXiv:2104.07991 [hep-th]].

\bibitem{merif}
J.~Evslin and H.~Liu,
``Quantum Reflective Kinks,''
[arXiv:2210.12725 [hep-th]].

\end{thebibliography}
\end{document}